\documentclass[twocolumn]{aastex701}

\usepackage{placeins}

\newcommand{\pandora}[0]{Pandora}
\newcommand{\kepler}[0]{Kepler}
\newcommand{\tess}[0]{TESS}
\newcommand{\hst}[0]{HST}
\newcommand{\jwst}[0]{JWST}

\newcommand{\vone}[1]{{#1}}

\shorttitle{Pandora Stellar Contamination Simulated Performance}
\shortauthors{Rackham et al.}
\received{February 17, 2026}
\revised{April 10, 2026}
\accepted{April 13, 2026}
\graphicspath{{./}{figures/}}

\begin{document}

\title{NASA's Pandora SmallSat Mission: Simulating the Impact of Stellar Photospheric Heterogeneity and Its Correction}

\author[0000-0002-3627-1676]{Benjamin V.\ Rackham}
\affiliation{Department of Earth, Atmospheric and Planetary Sciences, Massachusetts Institute of Technology, 77 Massachusetts Avenue, Cambridge, MA 02139, USA}
\affiliation{Kavli Institute for Astrophysics and Space Research, Massachusetts Institute of Technology, Cambridge, MA 02139, USA}
\email[show]{brackham@mit.edu}
\correspondingauthor{Benjamin V.\ Rackham}

\author[0000-0003-0971-1709]{Aishwarya R.\ Iyer}
\altaffiliation{NASA Postdoctoral Fellow}
\affiliation{NASA Goddard Space Flight Center, 8800 Greenbelt Road, Greenbelt, MD 20771, USA}
\email{aishwarya.iyer@nasa.gov}

\author[0000-0003-3714-5855]{D\'aniel Apai}
\affiliation{Steward Observatory, The University of Arizona, Tucson, AZ 85721, USA}
\affiliation{Lunar and Planetary Laboratory, The University of Arizona, Tucson, AZ 85721, USA}
\affiliation{James C. Wyant College of Optical Sciences, The University of Arizona, AZ 85721, USA}
\email{apai@arizona.edu}

\author[0000-0002-1052-6749]{Peter McGill}
\affiliation{Space Science Institute, Lawrence Livermore National Laboratory, 7000 East Ave., Livermore, CA 94550, USA}
\email{mcgill5@llnl.gov}

\author[0000-0003-4459-9054]{Yoav Rotman}
\affiliation{Space Science Institute, Lawrence Livermore National Laboratory, 7000 East Ave., Livermore, CA 94550, USA}
\affiliation{School of Earth and Space Exploration, Arizona State University, Tempe, AZ, USA}
\email{yrotman@asu.edu}

\author[0000-0001-8020-7121]{Knicole D. Col\'{o}n}
\affiliation{NASA Goddard Space Flight Center, 8800 Greenbelt Road, Greenbelt, MD 20771, USA}
\email{knicole.colon@nasa.gov}

\author[0000-0003-2528-3409]{Brett M. Morris}
\affiliation{Space Telescope Science Institute, 3700 San Martin Drive, Baltimore, MD 21218, USA}
\email{bmmorris@stsci.edu}

\author[0000-0002-0388-8004]{Emily A. Gilbert}
\affiliation{Caltech/IPAC-NASA Exoplanet Science Institute, Pasadena, CA 91125, USA}
\email{egilbert@caltech.edu}

\author[0000-0003-1309-2904]{Elisa V. Quintana}
\affiliation{NASA Goddard Space Flight Center, 8800 Greenbelt Road, Greenbelt, MD 20771, USA}
\email{elisa.quintana@nasa.gov}

\author[0000-0003-4206-5649]{Jessie L. Dotson}
\affiliation{NASA Ames Research Center, MS 245-6, Moffett Field, CA 94035, USA}
\email{jessie.dotson@nasa.gov}

\author[0000-0001-7139-2724]{Thomas Barclay}
\affiliation{NASA Goddard Space Flight Center, 8800 Greenbelt Road, Greenbelt, MD 20771, USA}
\email{thomas.barclay@nasa.gov}

\author{Pete Supsinskas}
\affiliation{Space Science Institute, Lawrence Livermore National Laboratory, 7000 East Ave., Livermore, CA 94550, USA}
\email{supsinskas1@llnl.gov}

\author{Jordan Karburn}
\affiliation{Space Science Institute, Lawrence Livermore National Laboratory, 7000 East Ave., Livermore, CA 94550, USA}
\email{karburn1@llnl.gov}

\author[0000-0002-3385-8391]{Christina Hedges}
\affiliation{University of Maryland, Baltimore County, 1000 Hilltop Circle, Baltimore, Maryland, United States}
\affiliation{NASA Goddard Space Flight Center, 8800 Greenbelt Road, Greenbelt, MD 20771, USA}
\email{christina.l.hedges@nasa.gov}

\author[0000-0002-5904-1865]{Jason F. Rowe}
\affiliation{Department of Physics and Astronomy, Bishops University, 2600 Rue College, Sherbrooke, QC J1M 1Z7, Canada}
\email{jason.rowe@ubishops.ca}

\author[0000-0002-5741-3047]{David R. Ciardi}
\affiliation{Caltech/IPAC-NASA Exoplanet Science Institute, Pasadena, CA 91125, USA}
\email{ciardi@ipac.caltech.edu}

\author[0000-0002-8035-4778]{Jessie L. Christiansen}
\affiliation{Caltech/IPAC-NASA Exoplanet Science Institute, Pasadena, CA 91125, USA}
\email{jessie.christiansen@caltech.edu}

\author[0000-0002-6276-1361]{Trevor O. Foote}
\altaffiliation{NASA Postdoctoral Fellow}
\affiliation{NASA Goddard Space Flight Center, 8800 Greenbelt Road, Greenbelt, MD 20771, USA}
\email{trevor.o.foote@nasa.gov}

\author[0000-0002-8963-8056]{Thomas P. Greene}
\affiliation{California Institute of Technology/IPAC, 1200 California Blvd, MC 100-22. Pasadena, CA 91125, USA}
\email{tgreene@ipac.caltech.edu}

\author[0000-0001-6541-0754]{Kelsey Hoffman}
\affiliation{Department of Physics and Astronomy, Bishops University, 2600 Rue College, Sherbrooke, QC J1M 1Z7, Canada}
\email{khoffman@ubishops.ca}

\author[0000-0002-5034-9476]{Rae Holcomb}
\altaffiliation{NASA Postdoctoral Fellow}
\affiliation{NASA Goddard Space Flight Center, 8800 Greenbelt Road, Greenbelt, MD 20771, USA}
\email{raeholcomb15@gmail.com}

\author[0000-0002-3239-5989]{Aurora Y. Kesseli}
\affiliation{Caltech/IPAC-NASA Exoplanet Science Institute, Pasadena, CA 91125, USA}
\email{aurorak@ipac.caltech.edu}

\author[0000-0001-9786-1031]{Veselin B. Kostov}
\affiliation{NASA Goddard Space Flight Center, 8800 Greenbelt Road, Greenbelt, MD 20771, USA}
\affiliation{SETI Institute, 189 Bernardo Ave, Suite 200, Mountain View, CA 94043, USA}
\email{veselin.b.kostov@nasa.gov}

\author[0000-0002-8507-1304]{Nikole K. Lewis}
\affiliation{Department of Astronomy and Carl Sagan Institute, Cornell University, 122 Sciences Drive, Ithaca, NY 14853, USA}
\email{nikole.lewis@cornell.edu}

\author[0000-0002-3783-5509]{James P. Mason}
\affiliation{Johns Hopkins University Applied Physics Laboratory, 11000 Johns Hopkins Rd, Laurel, MD 20723, USA}
\email{james.mason@jhuapl.edu}

\author[0000-0002-5982-566X]{Gregory Mosby Jr.}
\affiliation{NASA Goddard Space Flight Center, 8800 Greenbelt Road, Greenbelt, MD 20771, USA}
\email{gregory.mosby@nasa.gov}

\author[0000-0001-7106-4683]{Susan E. Mullally}
\affiliation{Space Telescope Science Institute, 3700 San Martin Drive, Baltimore, MD 21218, USA}
\email{smullally@stsci.edu}

\author[0000-0001-5347-7062]{Joshua E. Schlieder}
\affiliation{NASA Goddard Space Flight Center, 8800 Greenbelt Road, Greenbelt, MD 20771, USA}
\email{joshua.e.schlieder@nasa.gov}

\author[0000-0003-4241-7413]{Megan Weiner Mansfield}
\affiliation{Department of Astronomy, University of Maryland, College Park, MD 20742, USA}
\email{mwm@umd.edu}

\author[0000-0003-0156-4564]{Luis Welbanks}
\affiliation{School of Earth and Space Exploration, Arizona State University, Tempe, AZ, USA}
\email{luis.welbanks@asu.edu}

\author[0000-0002-1176-3391]{Allison Youngblood}
\affiliation{NASA Goddard Space Flight Center, 8800 Greenbelt Road, Greenbelt, MD 20771, USA}
\email{allison.a.youngblood@nasa.gov}

\begin{abstract}
Stellar photospheric heterogeneity is a dominant astrophysical systematic impacting exoplanet transmission spectroscopy.
NASA's Pandora SmallSat Mission is designed to address this challenge through contemporaneous visible-band photometry and near-infrared spectroscopy of exoplanet host stars.
Here, we present an end-to-end simulation study quantifying Pandora's ability to infer stellar photospheric properties and correct stellar contamination using out-of-transit observations.
We construct eight representative stellar activity scenarios and generate 160 simulated Pandora datasets, incorporating time-dependent stellar spectra, instrument response, and noise.
\vone{Given accurate models,} Bayesian retrievals of joint visible photometry (0.4--0.7\,$\micron$) and near-infrared spectroscopy (0.9--1.6\,$\micron$, $R{\approx}120$) recover photospheric temperatures with typical uncertainties of ${\approx}30$\,K, with no significant bias. 
Models with two spectral components (i.e., a quiescent photosphere and spots) are strongly favored in 95\% of cases; 
one-component models are preferred when true spot filling factors fall below a detection threshold of ${\approx}0.3\%$.
We propagate the true and inferred stellar parameters to compute true, inferred, and residual contamination signals under physically motivated spot geometries.
For simple spot distributions, contamination signals of $10^2$--$10^3$\,ppm are reduced to ${\lesssim}10$\,ppm---well below Pandora's expected transmission spectroscopy precision (30--100\,ppm).
For more complex spot distributions, geometric degeneracies limit deterministic corrections, leaving residual contamination at the $10^3$\,ppm level that must be mitigated using additional constraints, such as spot-crossing events and joint stellar--planetary retrievals of transmission spectra.
These results define regimes in which stellar contamination can be corrected from stellar observations alone and show how Pandora stellar observations can identify cases where additional information is required.
\end{abstract}



\section{Introduction}

Stellar spectral imprints arising from photospheric heterogeneity due to starspots and faculae have emerged as a fundamental limitation for exoplanet transmission spectroscopy.
Spatially inhomogeneous stellar surfaces imprint wavelength-dependent signals on transit depths that can bias inferred planetary radii, spectral slopes, and molecular abundances \citep{Pinhas2018, Iyer2020, Rackham2023}.
Termed the ``transit light source (TLS) effect,'' or more generally ``stellar contamination,'' these impacts are particularly severe for planets transiting active K- and M-dwarf hosts \citep{Rackham2018, Rackham2019}, which dominate current and future samples of small, potentially temperate exoplanets \citep{TJCI2024}.

In the era of JWST, where instrumental precision can reach tens of parts per million \citep[e.g.,][]{Rustamkulov2022, Rustamkulov2023_ERS, Ahrer2023, Alderson2023, Feinstein2023, JTECERST2023}, stellar contamination is no longer an occasional nuisance but a leading astrophysical systematic \citep[e.g.,][]{Lim2023, Moran2023, Ahrer2025, Espinoza2025_T1}.
Quantifying and mitigating the TLS effect is therefore essential to fully leverage high-quality transmission spectra \citep{Rackham2023, Rackham2024, Espinoza2026, Rathcke2025, Allen2026}. 
Effective mitigation is also critical for enabling potential future statistical studies of large samples of exo-Earth candidates (e.g., with sensitive transit-based surveys like the Nautilus concept; \citealt{Apai2019}).

NASA's Pandora SmallSat Mission \citep{Quintana2024, Barclay2025} was designed to directly address this challenge by enabling dedicated, contemporaneous characterization of exoplanet host stars alongside precise exoplanet atmospheric measurements.
Having recently launched in January 2026, \pandora{} provides a space-based platform in which a 0.45-m primary feeds two instruments simultaneously:
a visible-band photometer (VISDA; 0.4--0.7\,$\micron$) and a near-infrared (NIR) spectrograph (NIRDA; 0.9--1.6\,$\micron$, $R{\approx}120)$.
This configuration enables simultaneous broadband photometric monitoring and low-resolution NIR spectroscopy, optimized to constrain stellar surface heterogeneity and to support robust interpretation of exoplanetary transmission spectra.
Over its 1-year primary mission, \pandora{} will conduct an intensive monitoring campaign of 20 exoplanets transiting active K- and M-dwarf hosts, obtaining 10 transit observations per target \citep{Foote2023}.
Each visit spans 24 hours, enabling the collection of hours of out-of-transit baseline with multiwavelength measurements of the stellar photosphere contemporaneous with each planetary transit.

The central premise of \pandora{}'s stellar science program is that disk-integrated stellar spectra and photometry obtained outside of transit can be used to infer stellar surface properties at the time of transit, and thereby constrain or correct the imprint of stellar contamination on transmission spectra.
This observational strategy builds directly on recent results from the Hubble Space Telescope \citep[HST;][]{ZhangZhanbo2018, Wakeford2019, Garcia2022, Narrett2024} and JWST \citep[e.g.,][]{Lim2023, May2023, Moran2023, Ahrer2025, Rathcke2025}, which have shown that studying out-of-transit stellar spectra provides a critical consistency check on stellar photospheric properties inferred from in-transit data alone.
At the same time, it confronts a fundamental challenge in exoplanet characterization:
transmission spectra probe only a narrow, spatially resolved region of the projected stellar disk, while transit depths are necessarily normalized to the spectrum of the spatially unresolved, hemisphere-integrated star observed out of transit.
Bridging this mismatch requires accumulating the data and physical insights necessary to disentangle rotational modulation and evolving surface heterogeneities from integrated-light measurements, breaking the resulting spatial, spectral, and temporal degeneracies \citep{Rackham2023}.
However, the extent to which this approach succeeds depends on both the intrinsic properties of the star and the geometry of its surface heterogeneity relative to the transit chord.
A comprehensive, quantitative assessment of \pandora{}'s ability to recover stellar parameters and mitigate the TLS effect across realistic activity regimes is therefore essential.

Here, we present an end-to-end simulation analysis of \pandora{}'s expected performance in characterizing stellar photospheres and correcting stellar contamination in transmission spectra using out-of-transit observations alone.
We construct a suite of representative K- and M-dwarf activity scenarios spanning a range of rotation rates and spot morphologies.
For each scenario, we generate realistic stellar spectrophotometric time series, propagate them through a detailed \pandora{} instrument and noise model, and perform Bayesian retrievals on the 160 simulated datasets.
We then forward-model the true and inferred stellar parameters to quantify the true, inferred, and residual stellar contamination signals under a range of physically motivated spot geometries.
Our goal is to determine when the TLS effect can be effectively corrected using out-of-transit data, while also identifying regimes in which such corrections are intrinsically limited and quantifying the associated uncertainty.
A companion study \citep{Rotman2026} explores \pandora{}'s ability to constrain exoplanetary properties from in-transit data.

This paper is organized as follows. 
In \autoref{sec:targets}, we define the representative stellar scenarios and variability models used in our simulations.
\autoref{sec:data} describes the generation of simulated \pandora{} datasets, including instrument response and noise modeling.
In \autoref{sec:framework}, we present the stellar retrieval framework and model-selection approach.
\autoref{sec:results} summarizes the precision and accuracy with which stellar parameters are recovered.
In \autoref{sec:discussion}, we quantify stellar contamination signals, assess the effectiveness of contamination corrections, and discuss implications for the \pandora{} science goals.
We conclude in \autoref{sec:conclusions} with a summary of our findings.

\section{Representative Target Sample}
\label{sec:targets}

\begin{deluxetable*}{llccccccc}[!t]
\tablecaption{Simulated Stellar Targets and Spot Configurations\label{tab:scenarios}}
\tablehead{
\colhead{Scenario} & 
\colhead{Description} & 
\colhead{$T_{\mathrm{phot}}$ (K)} & 
\colhead{$T_{\mathrm{spot}}$ (K)} & 
\colhead{$P_{\mathrm{rot}}$ (d)} & 
\colhead{$R_\star$ ($R_\odot$)} & 
\colhead{$\log g$} & 
\colhead{$D_\star$ (pc)} & 
\colhead{$R_{\mathrm{spot}}$ ($\degr$)}
}
\startdata
\texttt{KFG} & K dwarf, fast rotator, giant spots  & 4700 & 3600 & 5  & 0.73 & 4.5 & 69.2 & 7 \\
\texttt{KFS} & K dwarf, fast rotator, solar-like spots  & 4700 & 3600 & 5  & 0.73 & 4.5 & 69.2 & 2 \\
\texttt{KSG} & K dwarf, slow rotator, giant spots  & 4700 & 3600 & 30 & 0.73 & 4.5 & 69.2 & 7 \\
\texttt{KSS} & K dwarf, slow rotator, solar-like spots  & 4700 & 3600 & 30 & 0.73 & 4.5 & 69.2 & 2 \\
\texttt{MFG} & M dwarf, fast rotator, giant spots  & 3400 & 3000 & 5  & 0.36 & 5.0 & 21.1 & 7 \\
\texttt{MFS} & M dwarf, fast rotator, solar-like spots  & 3400 & 3000 & 5  & 0.36 & 5.0 & 21.1 & 2 \\
\texttt{MSG} & M dwarf, slow rotator, giant spots  & 3400 & 3000 & 30 & 0.36 & 5.0 & 21.1 & 7 \\
\texttt{MSS} & M dwarf, slow rotator, solar-like spots  & 3400 & 3000 & 30 & 0.36 & 5.0 & 21.1 & 2 \\
\enddata
\tablecomments{Each scenario combines spectral type (K or M), rotation rate (fast or slow), and spot morphology (giant or solar-like).}
\end{deluxetable*}

\subsection{Definition of Simulated Target Stars}

We defined eight representative stellar scenarios to span a range of spectral types, rotation rates, and spot morphologies relevant to the \pandora{} target list\footnote{Available at \url{https://pandorasat.com/targets/}.}.
Although the target list may continue to evolve, these scenarios are designed to encompass the breadth of stellar properties expected among \pandora{} targets.
These representative targets are summarized in \autoref{tab:scenarios} and labeled with a concise shorthand, which is used hereafter.

To capture a range of stellar variability behavior, we include both K and M dwarfs with fast (5\,day) and slow (30\,day) rotation.
For each stellar type, we simulate photospheres with two typical spot sizes: ``solar-like'' spots with a radius of $2\,\degr$ and ``giant'' spots with radius of $7\,\degr$.
These values follow the boundary conditions used by \citet{Rackham2018}, approximately representing the largest sunspot groups observed on the Sun and the typical spot sizes accessible via molecular spectropolarimetric observations of active stars.

Stellar radii, surface gravities, distances, and temperatures are fixed for each spectral type.
Distances were chosen such that an unspotted star with the adopted photospheric temperature would have an apparent magnitude of $J = 9.0$, representative of typical \pandora{} targets.

\subsection{Variability Simulations}
\label{sec:variability_sims}

To define realistic time-dependent spot filling factors\footnote{We define ``spot covering fraction'' as the fraction of the total stellar photosphere covered by spots, and ``spot filling factor'' as the fraction of the projected stellar disk (visible hemisphere) covered by spots.} for use in our simulated datasets, we performed variability simulations using the \texttt{spotter} package \citep{Garcia2025}.
We adopted a $1\%$ peak-to-peak variability amplitude in the \pandora{} VISDA bandpass as a representative variability level, based on typical photometric modulation seen in active low-mass stars \citep{McQuillan2014, Newton2016, Rackham2018, Rackham2019}.
For context, Table~4 of \citet{Boyle2026} reports a mean variability amplitude of 0.3\% for stars in \pandora{}'s primary target list in the \tess{} bandpass, which probes redder wavelengths than VISDA and therefore generally shows lower amplitudes for the same spot coverage.
Rather than exploring detailed stellar surface structure, our goal with these simulations was to use a simplified set of spot sizes and distributions that, for a representative variability amplitude, generate a plausible range of time-dependent spot filling factors that span the parameter space relevant for \pandora{} observations.

We computed spot--photosphere contrasts using \texttt{pandora-sat}\footnote{\url{https://github.com/PandoraMission/pandora-sat}} with the adopted stellar parameters and VISDA bandpass sensitivity curve, yielding values of 0.28 for K dwarfs and 0.38 for M dwarfs.
Spots were modeled as circular features with fixed angular radii of either $2\degr$ (solar-like) or 7$\degr$ (giant), distributed isotropically over the stellar surface (i.e., uniformly over the surface area).
We assume an edge-on stellar inclination of $i_\star = 90^\circ$
as a representative case, as transiting exoplanet host stars are geometrically biased toward high stellar inclinations \citep{Hirano2014, Mazeh2015}.
Limb-darkening coefficients were obtained from the ExoCTK tool \citep{Bourque2021, Bourque2022} using a top-hat bandpass over 0.4--0.7\,$\micron$ and a quadratic law.

For each scenario, we first estimated the number of spots required to reproduce the adopted variability amplitude over a single rotation.
We ran 10 trial simulations per case, in which spots were added sequentially until reaching the $1\%$ variability.
Fixing this number of spots, we then simulated 100 spotted photospheres and recorded the average minimum and maximum projected spot filling factors over a full stellar rotation.
These values, denoted as $f_\mathrm{min}$ and $f_\mathrm{max}$, are reported in \autoref{tab:spot_coverage}.
Here, a value of $f_\mathrm{min} = 0$ does not imply an unspotted star but rather that, at some rotational phases, no spots are present in the visible hemisphere (e.g., existing spots reside on the far hemisphere).
Since the variability does not depend on the rotation period, fast and slow rotators modeled with the same spot morphology share identical spot distributions. 
We assume uniformly distributed spots for both fast- and slow-rotating M dwarfs given the lack of observational constraints on spot locations at these periods. 
We note, however, that even faster-rotating M dwarfs with $P_{\rm rot} < 1$ day have spots concentrated near the poles in Doppler imaging \citep{Barnes2015, Barnes2017}, in agreement with localized flare activity \citep{Ilin2024}, which can pose even larger challenges for transmission spectroscopy.

\begin{figure*}[htbp]
    \centering
    \includegraphics[width=\linewidth]{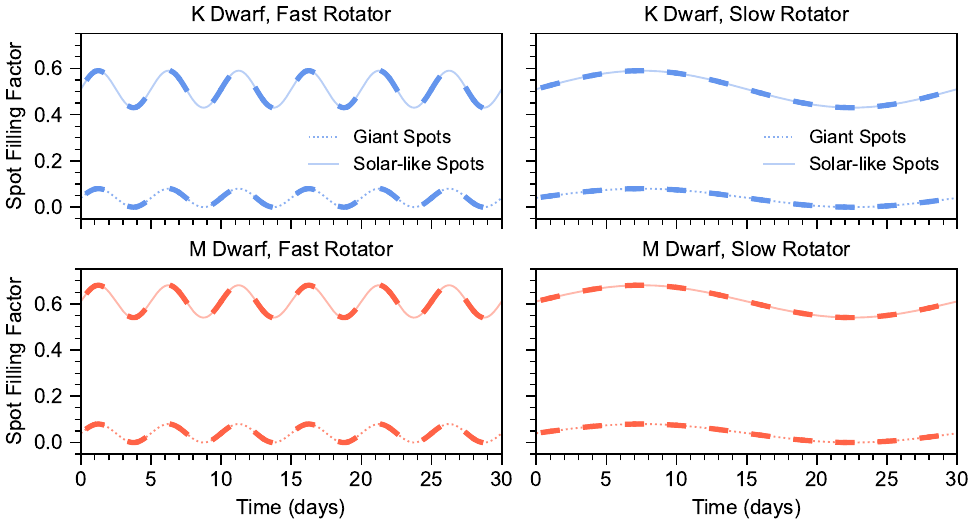}
    \caption{
        Simulated time-dependent spot filling factors for each stellar scenario.
        Each panel shows sinusoidal variations in the projected spot filling factor over a full 30-day observing window, using the adopted stellar rotation period (5 or 30\,d) and the $f_{\mathrm{min}}$ and $f_{\mathrm{max}}$ values from \autoref{tab:spot_coverage}.
        Thicker line segments show the times of the simulated \pandora{} observations.
    }
    \label{fig:fspot}
\end{figure*}

We used the $f_\mathrm{min}$ and $f_\mathrm{max}$ values to define a time-dependent spot filling factor for each scenario, which varies sinusoidally between $f_\mathrm{min}$ and $f_\mathrm{max}$ over the stellar rotation period (\autoref{fig:fspot}).
These simulations are intentionally simplified, as our goal is to assess \pandora{}'s performance under controlled, physically motivated stellar variability, rather than to reproduce a detailed stellar surface structure.
The adopted simplifications result in spot filling factors that span a broad range of 0--68\% (\autoref{tab:spot_coverage}), covering plausible activity levels that may be present in \pandora{} observations.
Importantly, the simulated variability is used only to define the instantaneous filling factor at each epoch;
the specific temporal form (e.g., sinusoidal evolution) is not imposed in the retrieval framework, which treats each visit independently.
The resulting spot filling factor time series are used to generate the simulated \pandora{} datasets, as described in \autoref{sec:data}.

\begin{deluxetable}{lccc}
\tablecaption{Summary of Variability Simulations\label{tab:spot_coverage}}
\tablehead{
\colhead{Scenarios} & 
\colhead{$n_\mathrm{spot}$} & 
\colhead{$f_{\mathrm{min}}$} & 
\colhead{$f_{\mathrm{max}}$}
}
\startdata
\texttt{KFG, KSG} & 1   & 0.00 & 0.08 \\
\texttt{KFS, KSS} & 73  & 0.43 & 0.59 \\
\texttt{MFG, MSG} & 2   & 0.00 & 0.08 \\
\texttt{MFS, MSS} & 102 & 0.54 & 0.68 \\
\enddata
\tablecomments{
Each entry combines the fast and slow rotator cases for a given spectral type and spot morphology, which share the same spot distribution parameters.
}
\end{deluxetable}

\section{Simulated Pandora Datasets}
\label{sec:data}

\begin{figure*}[htbp]
    \centering
    \includegraphics[width=\textwidth]{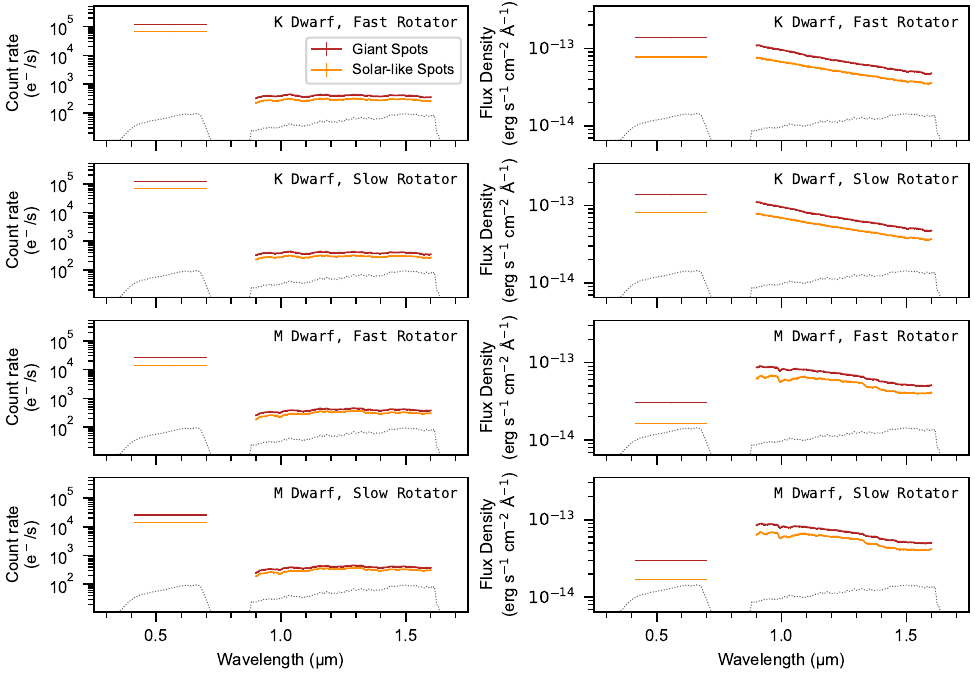}
    \caption{Representative simulated \pandora{} spectra for each stellar scenario.
    Each row corresponds to a stellar scenario, while the left and right columns show count rate and flux density, respectively.
    In each panel, a representative 12-hr binned spectrum is shown for both giant-spot (red) and solar-like spot (orange) morphologies.
    Vertical error bars, which are generally smaller than the line width, denote 12-hr measurement uncertainties. 
    The dotted black curves indicate the sensitivity of the VISDA and NIRDA instruments, shown on an arbitrary scale.
    }
    \label{fig:count_rates}
\end{figure*}

To test how effectively \pandora{} observations can recover stellar photospheric properties in the presence of realistic stellar heterogeneity, we generated a suite of synthetic datasets for each variability scenario using the time series of spot filling factors defined in our variability analysis (\autoref{sec:variability_sims}).
All component spectra (spots and photosphere) are drawn from the PHOENIX ACES stellar atmosphere grid \citep{husser2013} using \texttt{speclib} \citep{speclib2023, speclib2025}.
We use solar abundances ([Fe/H]) for all spectral components, allowing us to isolate the impact of surface heterogeneity.

\subsection{True Model Parameters}

For each scenario, we used the ``true'' set of stellar parameters given in \autoref{tab:scenarios}, including the photospheric temperature ($T_\mathrm{phot}$), spot temperature ($T_\mathrm{spot}$), surface gravity ($\log g$), stellar radius ($R_\star$), and distance ($D_\star$).
At each simulated time step, we constructed the total emergent stellar spectrum as a linear combination of the photosphere and spot components, weighted by the time-dependent spot filling factor (\autoref{fig:fspot}):
\begin{equation}
F_{\star}(t) = \big[ 1 - f_{\mathrm{spot}}(t) \big] F_{\mathrm{phot}} + f_{\mathrm{spot}}(t) F_{\mathrm{spot}}.
\end{equation}
The resulting $F_\star$ gives the emergent flux at the stellar surface. 
We scale this by $(R_\star / D_\star)^2$ to compute the spectral flux density at Earth ($\mathrm{erg\,s^{-1}\,cm^{-2}\,\mathring{A}^{-1}}$), consistent with the flux level \pandora{} would observe.

\subsection{Temporal Sampling}

For each scenario, we generated simulated observations at time points reflecting a realistic \pandora{} visit strategy.
Each synthetic time series spans 10 visits, with each visit covering a 24-hr window around transit.
We assumed a 3-day orbital period for the exoplanet and exclude from our analysis a 2-hr window around each transit midtime, as we are interested here only in the constraints \pandora{} enables for stellar photospheres.

Within each visit, we sampled the NIRDA spectra at a cadence of 33.6\,s and the VISDA photometry at a cadence of 10\,s.
These cadences reflect \pandora{}'s anticipated sampling for a $J{=}9.0$ target.
The final observation times are shown by the thicker line segments in \autoref{fig:fspot}.

\subsection{Instrument Response and Noise Model}

We used \texttt{pandora-sat} \citep{Hedges2024} to model the response of \pandora{}'s VISDA and NIRDA detectors to each disk-integrated spectrum.
For VISDA, we computed a single broadband flux weighted by the visible transmission curve.
For NIRDA, we convolved each spectrum with a Gaussian kernel to simulate the spectral resolving power of NIRDA ($R = 120$ at $1.3\,\micron$), multiplied by the instrument's sensitivity curve, and then resampled the spectrum using the wavelength-to-pixel mapping provided by \texttt{pandora-sat}.
For both VISDA and NIRDA, this process yielded the instrument's response in units of $e^-\mathrm{s}^{-1}$.

We then applied a realistic noise model to each simulated count rate.
The noise model includes photon noise (modeled as Poisson noise based on the total electron counts), as well as contributions from read noise, dark current, and background levels using the instrument performance parameters implemented in \texttt{pandora-sat} \citep{Hedges2024}.
All noise terms are combined in quadrature.
In the 12-hr bins, the median VISDA fractional uncertainty is ${\approx}30$\,ppm and the median NIRDA fractional uncertainty is ${\approx}5 \times 10^3$\,ppm per spectral bin.

\subsection{Data Products}

\autoref{fig:count_rates} shows example binned count rates for representative scenarios across all combinations of spectral type, rotation rate, and spot morphology.
In addition to modeling count rates, we converted the simulated measurements back into spectral flux densities using precomputed reference count rates corresponding to a flat input spectrum.
For downstream analysis, the time-series are additionally binned into 12-hr windows before and after each transit, excluding $\pm1$\,hr around the transit center and propagating uncertainties, assuming statistically independent samples.
This yields 160 simulated datasets for this analysis: eight stellar scenarios evaluated at 20 time indices each.

The simulated datasets used in this analysis are available on Zenodo,\footnote{\dataset[doi:10.5281/zenodo.18624584]{\doi{10.5281/zenodo.18624584}}} including native-cadence data as well as 12-hr binned time series.
Each simulated dataset includes noiseless and simulated (noisy) \pandora{} count-rate time series for VISDA and NIRDA (e$^{-}$\,s$^{-1}$), with associated $1\sigma$ uncertainties and corresponding spectral flux densities (erg\,s$^{-1}$\,cm$^{-2}$\,\AA$^{-1}$).
The metadata also includes the adopted stellar parameters and spot filling factor time series.
All products are provided in HDF5 format.

\section{Inference Framework}
\label{sec:framework}

Using these simulated datasets, we perform retrievals to test how effectively \pandora{} can constrain stellar photospheric properties in the presence of realistic spot-induced variability.
We perform these retrievals independently for each simulated time bin, fitting the VISDA and NIRDA measurements jointly.

\subsection{Forward Model}

Our forward model again uses the PHOENIX ACES grid to generate both photosphere and spot spectra.
We note that this setup assumes perfect model fidelity, which is a challenge for K and M dwarfs \citep{Iyer2020, Rackham2024}; the impact of relaxing this assumption is explored in \autoref{sec:fidelity}.
We do not interpolate between the models in the grid when evaluating stellar spectra.
Instead, for each sampled temperature, we adopt the nearest grid model.
This approach avoids interpolation-induced biases that can arise in high signal-to-noise retrievals \citep{Czekala2015, Rackham2024}.

We test both one- and two-component models for each scenario.
One-component retrievals assume a single homogeneous photosphere, while two-component retrievals allow for a composite spectrum with contributions from the quiescent photosphere and a cooler spotted region.
In both cases, the spectrum is scaled by $(R_\star / D_\star)^2$ to convert surface fluxes to flux densities at Earth.

To ensure consistency with the simulated datasets, the forward model also uses detector count units.
For each model evaluation, the PHOENIX surface flux spectrum is interpolated onto a high-resolution wavelength grid and convolved with a  Gaussian line-spread function corresponding to NIRDA's resolving power ($R = 120$ at $1.3\,\micron$).
The convolved spectrum is then integrated against the \texttt{pandora-sat} instrument sensitivity curves to generate detector count rates.
This procedure yields a single band-integrated count rate per time bin for VISDA and a vector of per-bin count rates over the NIRDA bandpass (0.9--1.6\,$\micron$).
Finally, the count rates are converted back into flux densities using reference counts computed for a flat spectrum, matching the calibration used in the data simulation.

\subsection{Parameterization and Priors}

\autoref{tab:priors} lists the scenario-specific priors we apply to reflect the expected range of stellar properties.
In the one-component model, we sample $T_\mathrm{phot}$ and $R_\star$.
In the two-component model, we sample $T_\mathrm{phot}$, $\Delta T$, $f_\mathrm{spot}$, and $R_\star$, where $T_\mathrm{spot} = T_\mathrm{phot} - \Delta T$.
The adopted priors are intentionally broad and conservative, allowing us to assess the information content of the \pandora{} observations themselves while avoiding the introduction of external constraints into the analysis.

To ensure sampled temperatures remain within the domain of the PHOENIX model grid, we draw $\Delta T$ from a uniform distribution on $[0, T_\mathrm{phot} - T_\mathrm{min}]$, where $T_\mathrm{min}$ is the lower bound of the sampled temperature space.
This construction ensures that $T_\mathrm{spot}$ always satisfies $T_\mathrm{min} \le T_\mathrm{spot} \le T_\mathrm{phot}$.
Similarly, $f_\mathrm{spot}$ is drawn from a uniform prior on $(\epsilon, 1 - \epsilon)$ with $\epsilon = 10^{-6}$, effectively sampling all physical valid spot coverages while avoiding exactly zero-area components.

As the data constrain only the flux-scaling factor of $(R_\star / D_\star)^2$, we fix $D_\star$ and fit only $R_\star$.
We adopt the true values of $D_\star$ (\autoref{tab:scenarios}), and place truncated normal priors on $R_\star$.
This choice reduces the dimensionality of the parameter space and speeds up convergence, and it is appropriate for our purposes because we are not interested in independent constraints on $R_\star$ and $D_\star$ but instead to evaluate the recovery of photospheric and spot properties.
The inferred $R_\star$ parameter therefore functions primarily to scale the flux.
Finally, we fix $\mathrm{[Fe/H]}$ and $\log g$ to the true values and do not marginalize over these quantities in the retrievals.

\subsection{Sampling and Likelihood}

We use \texttt{UltraNest}'s \citep{Buchner2021} reactive nested sampling framework to perform the parameter estimation.
As the step sampler, we adopt \texttt{UltraNest}'s \texttt{SliceSampler} because it efficiently explores strongly correlated posteriors without requiring tuned proposal scales.
In each sampling, we set the number of steps to 10 times the number of parameters and adopt a minimum of 1024 live points to ensure robust exploration of the posterior volume and stable evidence estimation.
Sampling is terminated when $\Delta \ln \mathcal{Z} < 0.5$, ensuring that the uncertainty in the estimated Bayesian evidence is small compared to the evidence differences used for model selection.

We adopt a Gaussian likelihood assuming uncorrelated noise to compare the forward model to the observed fluxes and their uncertainties.
The likelihood is evaluated on the concatenated data vector comprising the VISDA flux density point and the NIRDA flux density spectrum, with independent (diagonal) uncertainties.

\begin{deluxetable}{llcc}
\tabletypesize{\footnotesize}  
\tablecaption{Priors Used in Retrievals\label{tab:priors}}
\tablehead{
\colhead{Model} &
\colhead{Parameter} &
\colhead{K Dwarfs} &
\colhead{M Dwarfs}
}
\startdata
1-comp & $T_{\mathrm{phot}}$\,(K) & $\mathcal{U}$(3000, 5000) & $\mathcal{U}$(2300, 4300) \\
2-comp & $T_{\mathrm{phot}}$\,(K) & $\mathcal{U}$(3000, 5000) & $\mathcal{U}$(2300, 4300) \\
2-comp & $\Delta T$\,(K) & $\mathcal{U}(0,\,T_{\mathrm{phot}}-T_{\min})$ & $\mathcal{U}(0,\,T_{\mathrm{phot}}-T_{\min})$ \\
2-comp & $f_{\mathrm{spot}}$ & $\mathcal{U}(\epsilon, 1-\epsilon)$ & $\mathcal{U}(\epsilon, 1-\epsilon)$ \\
All & $R_\star$ ($R_\odot$) & $\mathcal{N}(0.73,\,0.02)$ & $\mathcal{N}(0.36,\,0.01)$ \\
All & $[{\rm Fe}/{\rm H}]$ & 0.0 (fixed) & 0.0 (fixed) \\
All & $\log g$ & 4.5 (fixed) & 5.0 (fixed) \\
\enddata
\tablecomments{
One- and two-component models are abbreviated as ``1-comp'' and ``2-comp,'' respectively.
$\mathcal{U}(a,b)$ denotes a uniform prior between $a$ and $b$, and $\mathcal{N}(\mu,\sigma)$ denotes a normal distribution with mean $\mu$ and standard deviation $\sigma$.
The stellar radius prior is a truncated normal distribution, with $\sigma$ equal to 3\% of the nominal value and bounds $[0.6,\,0.9]\,R_\odot$ for K dwarfs and $[0.25,\,0.5]\,R_\odot$ for M dwarfs to improve sampling efficiency.
In the two-component model, the spot temperature is defined as $T_{\mathrm{spot}} = T_{\mathrm{phot}} - \Delta T$. The upper bound on $\Delta T$ enforces $T_{\mathrm{spot}} \ge T_{\min}$, where $T_{\min}=3000$~K for K dwarfs and $T_{\min}=2300$~K for M dwarfs.
}
\end{deluxetable}

\subsection{Model Selection}

For each scenario and time bin, we select the preferred model between the one- and two-component fits using the Bayesian evidence, which is estimated directly by \texttt{UltraNest} as part of the nested sampling process.
We consider differences in log-evidence of $\Delta \ln \mathcal{Z} > 11$ to indicate strong support for the more complex model, corresponding to odds greater than $\sim 6 \times 10^{4}\!:\!1$ in its favor.
Following common practice for interpreting Bayesian evidence \citep{Trotta2008, Benneke2013}, this corresponds to approximately a $5\sigma$ result, though we note that these Bayesian-to-frequentist conversions yield optimistic values of $\sigma$ \citep{Kipping2025} and should be more precisely understood as upper limits on the true significance.

\section{Results}
\label{sec:results}

Here, we present the results of our simulated retrievals, focusing on how well Pandora-like observations can constrain stellar photospheric and spot properties across a range of representative scenarios \vone{when model fidelity is high}.
We first examine what level of model complexity (one or two components) is preferred for each scenario.
We then quantify the typical precision and accuracy of the retrieved stellar parameters.

\subsection{Model Preference}

Across the full ensemble of 160 simulated datasets, the two-component model is strongly preferred in 152 cases (95\%) at $\Delta \ln \mathcal{Z} \gg 11$ confidence.
The remaining eight datasets show a weak to moderate preference for the one-component model, with $\Delta \ln \mathcal{Z} \simeq 3.1{-}3.6$ ($\leq 2.5{-}2.7\sigma$; \autoref{tab:exceptions}).
These cases are confined exclusively to the giant-spot scenarios and occur at time indices when the true spot filling factors were $\ll 1\%$.
We discuss this result in \autoref{sec:interpretation}.

\begin{deluxetable*}{lrrrr}
\tabletypesize{\footnotesize}
\tablecaption{Datasets Where the One-component Model is Preferred \label{tab:exceptions}}
\tablehead{
\colhead{Scenario} & \colhead{Time Index} & \colhead{$\Delta \ln \mathcal{Z}$} & \colhead{$\sigma_\mathrm{max}$} & \colhead{True $f_{\rm spot}$ (ppm)}
}
\startdata
\texttt{KFG} & 2  & 3.4 & 2.6 & 564 \\
\texttt{KFG} & 12 & 3.5 & 2.7 & 564 \\
\texttt{KSG} & 14 & 3.6 & 2.7& 535 \\
\texttt{KSG} & 15 & 3.5 & 2.7 & 61 \\
\texttt{MFG} & 2  & 3.2 & 2.5 & 564 \\
\texttt{MFG} & 12 & 3.3 & 2.6 & 564 \\
\texttt{MSG} & 14 & 3.1 & 2.5 & 535 \\
\texttt{MSG} & 15 & 3.2 & 2.5 & 61 \\
\enddata
\tablecomments{
Here $\Delta \ln \mathcal{Z} \equiv \ln \mathcal{Z}_{\rm best} - \ln \mathcal{Z}_{\rm second}$, where the best model is the one-component fit in all listed cases.
The quoted $\sigma_\mathrm{max}$ values represent upper limits on the equivalent frequentist significance \citep{Kipping2025}.
}
\end{deluxetable*}

\subsection{Inference Precision and Accuracy}

\begin{figure*}[p]
    \centering
    \includegraphics[width=\textwidth]{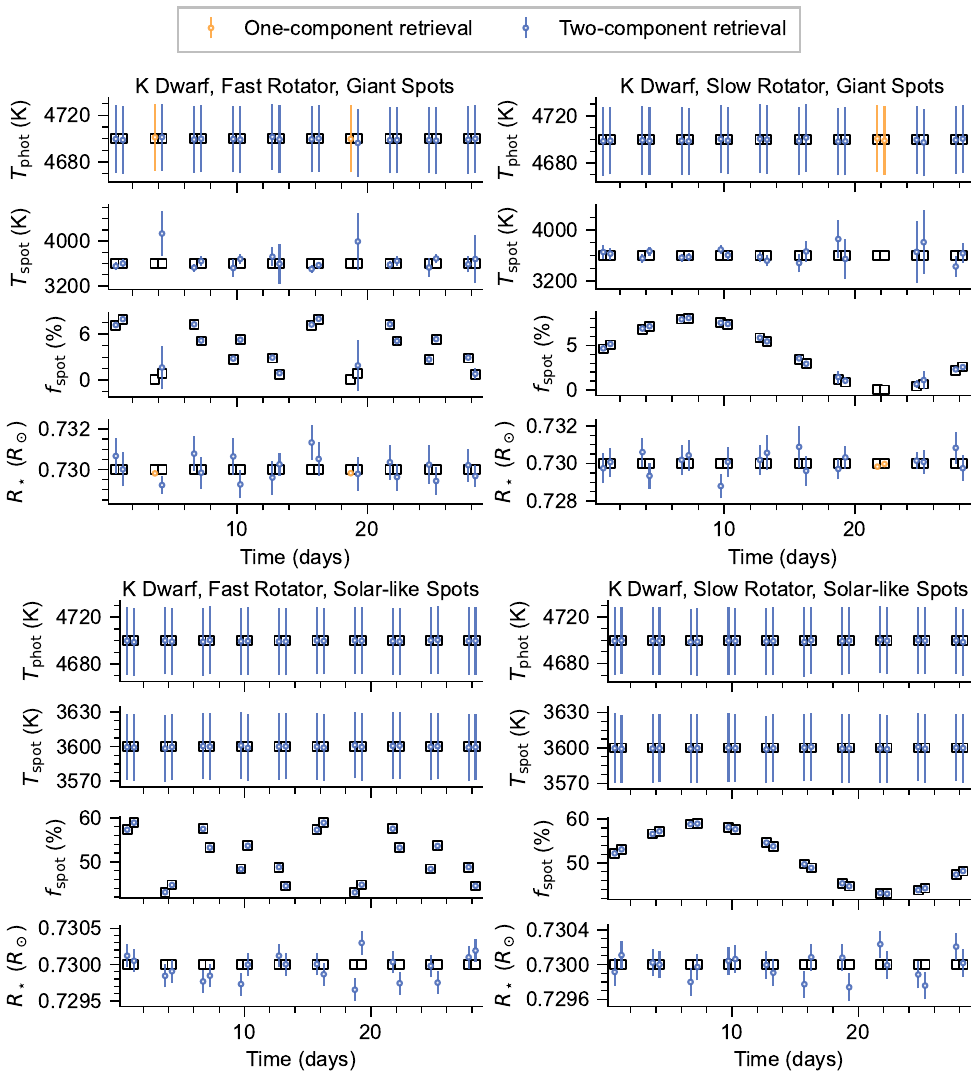}
    \caption{
    Time series of inferred stellar parameters from fits to simulated \pandora{} observations of K-dwarf targets under four variability scenarios.
    Panels show the recovered photospheric temperature ($T\mathrm{phot}$), spot temperature ($T_\mathrm{spot}$), spot filling factor ($f_\mathrm{spot}$), and stellar radius ($R_\star$) as a function of time.
    Each point corresponds to a pre- or post-transit stellar spectrum from an individual 24-hr visit.
    Black boxes indicate the input (true) parameter values.
    Blue points show the posterior means from the two-component retrievals, with error bars denoting the 68\% credible intervals.
    Orange points indicate datasets for which the one-component model is  preferred, for which $T_\mathrm{spot}$ and $f_\mathrm{spot}$ are not defined.
    }
    \label{fig:inferences_K}
\end{figure*}

\begin{figure*}[p]
    \centering
    \includegraphics[width=\textwidth]{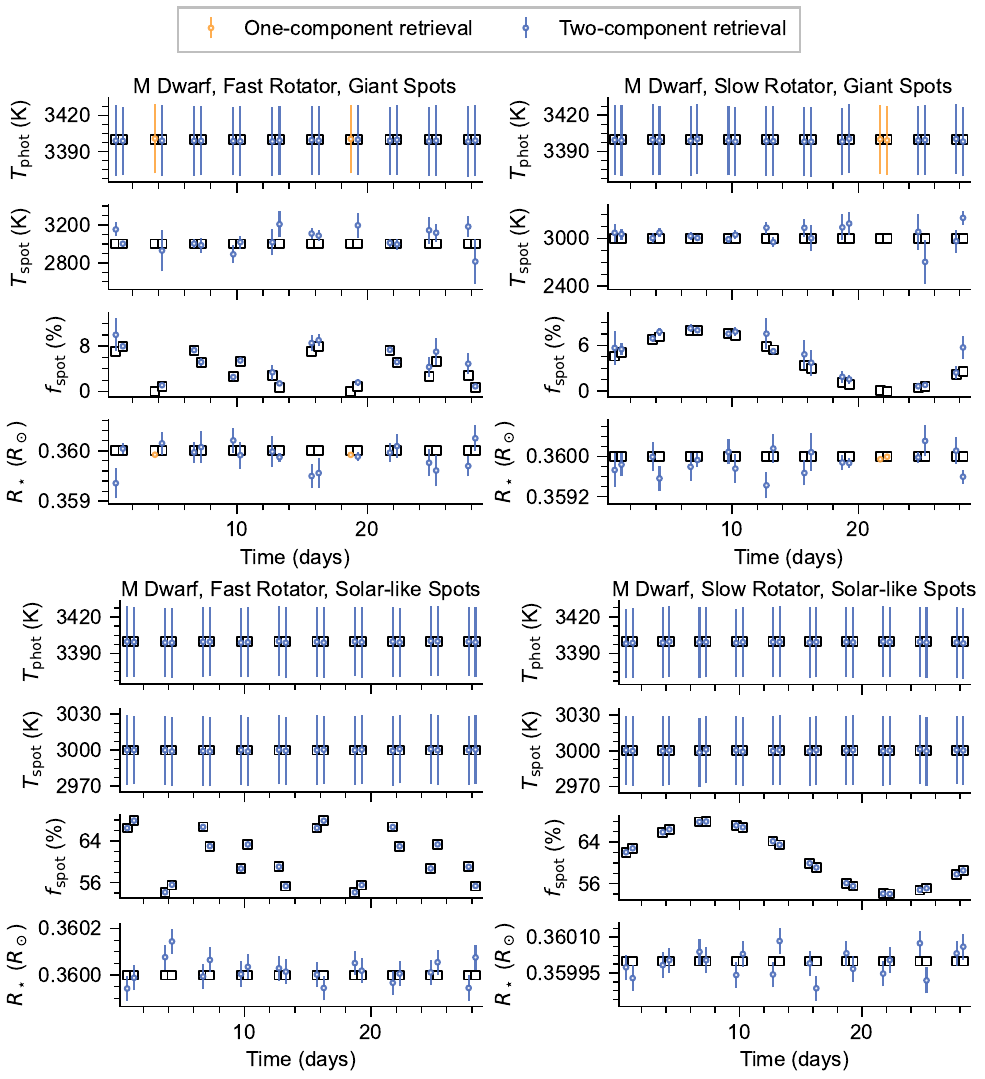}
    \caption{
    Time series of inferred stellar parameters from fits to simulated \pandora{} observations of M-dwarf targets under four variability scenarios.
    The figure elements are the same as in \autoref{fig:inferences_K}.
    }
    \label{fig:inferences_M}
\end{figure*}

\begin{deluxetable*}{llrrrr}
\tablecaption{Precision and Accuracy of Inferred Stellar Parameters \label{tab:precision_accuracy}}
\tablehead{
\colhead{Scenario} &
\colhead{Parameter} &
\colhead{Mean($\sigma$)} &
\colhead{SD($\sigma$)} &
\colhead{Mean($b$)/$\sigma$} &
\colhead{RMS($b$)/$\sigma$}
}
\startdata
\texttt{KFG} & $T_\mathrm{phot}$ (K) & $29$ & $0.30$ & $-0.024$ & $0.045$ \\
\texttt{KFG} & $T_\mathrm{spot}$ (K) & $174$ & $147$ & $0.051$ & $0.75$ \\
\texttt{KFG} & $f_\mathrm{spot}$ (\%) & $0.52$ & $0.96$ & $0.49$ & $0.61$ \\
\texttt{KFG} & $R_\star$ ($R_\odot$) & $7.7\times 10^{-4}$ & $1.4\times 10^{-4}$ & $0.014$ & $0.79$ \\
\hline
\texttt{KFS} & $T_\mathrm{phot}$ (K) & $29$ & $0.19$ & $-0.011$ & $0.020$ \\
\texttt{KFS} & $T_\mathrm{spot}$ (K) & $29$ & $0.22$ & $-4.7\times 10^{-3}$ & $0.025$ \\
\texttt{KFS} & $f_\mathrm{spot}$ (\%) & $0.028$ & $2.7\times 10^{-3}$ & $-0.32$ & $1.1$ \\
\texttt{KFS} & $R_\star$ ($R_\odot$) & $1.6\times 10^{-4}$ & $2.0\times 10^{-6}$ & $-0.31$ & $1.1$ \\
\hline
\texttt{KSG} & $T_\mathrm{phot}$ (K) & $29$ & $0.24$ & $-0.027$ & $0.044$ \\
\texttt{KSG} & $T_\mathrm{spot}$ (K) & $167$ & $140$ & $0.050$ & $0.67$ \\
\texttt{KSG} & $f_\mathrm{spot}$ (\%) & $0.22$ & $0.27$ & $0.42$ & $0.66$ \\
\texttt{KSG} & $R_\star$ ($R_\odot$) & $7.5\times 10^{-4}$ & $1.5\times 10^{-4}$ & $0.049$ & $0.68$ \\
\hline
\texttt{KSS} & $T_\mathrm{phot}$ (K) & $29$ & $0.29$ & $-0.023$ & $0.030$ \\
\texttt{KSS} & $T_\mathrm{spot}$ (K) & $29$ & $0.28$ & $-6.5\times 10^{-3}$ & $0.020$ \\
\texttt{KSS} & $f_\mathrm{spot}$ (\%) & $0.036$ & $0.035$ & $-0.064$ & $0.82$ \\
\texttt{KSS} & $R_\star$ ($R_\odot$) & $1.6\times 10^{-4}$ & $1.8\times 10^{-6}$ & $-0.13$ & $0.87$ \\
\hline
\texttt{MFG} & $T_\mathrm{phot}$ (K) & $29$ & $0.34$ & $-0.034$ & $0.037$ \\
\texttt{MFG} & $T_\mathrm{spot}$ (K) & $101$ & $58$ & $0.60$ & $1.1$ \\
\texttt{MFG} & $f_\mathrm{spot}$ (\%) & $0.96$ & $0.82$ & $0.61$ & $0.77$ \\
\texttt{MFG} & $R_\star$ ($R_\odot$) & $2.3\times 10^{-4}$ & $7.2\times 10^{-5}$ & $-0.53$ & $1.1$ \\
\hline
\texttt{MFS} & $T_\mathrm{phot}$ (K) & $29$ & $0.23$ & $-0.021$ & $0.026$ \\
\texttt{MFS} & $T_\mathrm{spot}$ (K) & $29$ & $0.23$ & $-2.7\times 10^{-3}$ & $0.020$ \\
\texttt{MFS} & $f_\mathrm{spot}$ (\%) & $0.022$ & $1.6\times 10^{-3}$ & $0.36$ & $1.00$ \\
\texttt{MFS} & $R_\star$ ($R_\odot$) & $5.4\times 10^{-5}$ & $5.4\times 10^{-7}$ & $0.34$ & $0.99$ \\
\hline
\texttt{MSG} & $T_\mathrm{phot}$ (K) & $29$ & $0.27$ & $-0.032$ & $0.042$ \\
\texttt{MSG} & $T_\mathrm{spot}$ (K) & $108$ & $67$ & $0.57$ & $1.1$ \\
\texttt{MSG} & $f_\mathrm{spot}$ (\%) & $0.86$ & $0.68$ & $0.60$ & $0.87$ \\
\texttt{MSG} & $R_\star$ ($R_\odot$) & $2.4\times 10^{-4}$ & $7.9\times 10^{-5}$ & $-0.65$ & $1.2$ \\
\hline
\texttt{MSS} & $T_\mathrm{phot}$ (K) & $29$ & $0.28$ & $-0.029$ & $0.033$ \\
\texttt{MSS} & $T_\mathrm{spot}$ (K) & $29$ & $0.23$ & $2.9\times 10^{-3}$ & $0.022$ \\
\texttt{MSS} & $f_\mathrm{spot}$ (\%) & $0.022$ & $1.5\times 10^{-3}$ & $-0.19$ & $0.99$ \\
\texttt{MSS} & $R_\star$ ($R_\odot$) & $5.4\times 10^{-5}$ & $6.0\times 10^{-7}$ & $-0.14$ & $1.00$ \\
\hline
All & $T_\mathrm{phot}$ (K) & $29$ & $0.27$ & $-0.025$ & $0.035$ \\
All & $T_\mathrm{spot}$ (K) & $80$ & $95$ & $0.15$ & $0.64$ \\
All & $f_\mathrm{spot}$ (\%) & $0.32$ & $0.61$ & $0.22$ & $0.87$ \\
All & $R_\star$ ($R_\odot$) & $2.9\times 10^{-4}$ & $2.8\times 10^{-4}$ & $-0.16$ & $0.98$ \\
\enddata
\tablecomments{
For each scenario and parameter, Mean($\sigma$) and SD($\sigma$) summarize the mean and standard deviation of the distribution of posterior uncertainties across time indices and are reported in the native units of the parameter. 
Mean($b$)/$\sigma$ and RMS($b$)/$\sigma$ summarize the mean signed and root-mean-square normalized residuals $b \equiv (\hat{\theta}-\theta_\mathrm{true})/\sigma$ across time indices and are therefore in units of $\sigma$ (dimensionless). 
Only time indices for which the two-component model is preferred are included in these summaries. 
The ``All'' rows summarize results pooled across all scenarios.
We note that these rows combine results across heterogeneous scenarios, and for parameters such as $T_\mathrm{spot}$ and $f_\mathrm{spot}$ this can average over a multi-modal population (e.g., the $\sim$30\,K and $\sim$100--170\,K $T_\mathrm{spot}$ regimes evident in the scenario-level Mean($\sigma$) values), such that the reported precision is not representative of any single physical case.
}
\end{deluxetable*}

Figures~\ref{fig:inferences_K} and \ref{fig:inferences_M} summarize the inferred stellar parameters for the K- and M-dwarf targets, respectively.
Each figure shows the posterior means and 68\% credible intervals for the parameters as a function of time, with each point corresponding to a pre- or post-transit stellar spectrum from an individual 24-hr visit. In all cases, the inferred parameters fluctuate about the input values, with no evidence for systematic offsets.
The scatter is consistent with the quoted uncertainties.

For the K-dwarf targets (\autoref{fig:inferences_K}), the photospheric temperature is well constrained in all variability scenarios.
Spot temperatures are recovered with a similar level of precision in the solar-like spot scenarios, while the giant-spot scenarios show reduced precision when the spot filling factor is $\lesssim1\%$.
In these same cases, the inferred spot filling factors show increased scatter.

The stellar radius parameter shows a similar level of scatter and uncertainty across all K-dwarf scenarios.
In these simulations, the stellar distance is fixed, so $R_\star$ primarily captures the precision with which the absolute flux normalization of the spectrum, $(R_\star / D_\star)^2$, is recovered.
The resulting tight constraints therefore reflect the information content of the overall flux level rather than a standalone measurement of $R_\star$.
In realistic applications, where distance would also carry uncertainty or be jointly inferred, the true stellar radius uncertainty would be correspondingly larger.

The M-dwarf results (\autoref{fig:inferences_M}) show qualitatively similar behavior.
The photospheric temperature remains well constrained across all scenarios and epochs, while the precision of the spot temperature and filling factor depends on the filling factor.
With relatively high values of $f_\mathrm{spot}$, solar-like spot scenarios yield tighter constraints on spot properties, whereas giant-spot scenarios with lower values of $f_\mathrm{spot}$ show reduced precision and increased scatter.
As with the K-dwarf results, the inferred values of $R_\star$, which again functions as a proxy for the flux normalization, show comparable scatter across all scenarios.

\autoref{tab:precision_accuracy} provides a quantitative summary of the precision and accuracy of the inferred stellar parameters across all scenarios, considering only epochs for which the two-component model is preferred.
The uncertainty on the photospheric temperature is remarkably uniform, with a mean value of 29\,K in every scenario and a pooled value of $29 \pm 0.27$\,K across all cases.
As the stellar spectra are evaluated on a discrete temperature grid with 100\,K spacing and no interpolation between models, the uncertainty on $T_\mathrm{phot}$ reflects how strongly the data discriminate between adjacent grid models.
This typical uncertainty implies that the likelihood strongly favors the best-fitting grid point over its nearest neighbors:
under a Gaussian approximation, this corresponds to a log-likelihood separation of $\Delta \ln \mathcal{L} \sim (100/29)^2 \approx 12$, or roughly a $3{-}4\sigma$ distinction between adjacent models.

Uncertainties on the spot temperature are generally larger and more variable.
They range from ${\sim}29$\,K in the solar-like spot scenarios to ${\sim}100{-}170$\,K in the giant-spot scenarios, with a pooled value of $80 \pm 95$\,K.
A similar pattern is evident for the spot filling factor, for which uncertainties are small in the solar-like spot cases and substantially larger in the giant-spot cases, yielding a pooled uncertainty of $0.32 \pm 0.61\%$.
In general, uncertainties on $T_\mathrm{spot}$ and $f_\mathrm{spot}$ are smaller when $f_\mathrm{spot}$ is large and increase as $f_\mathrm{spot}$ decreases (i.e., when spots contribute less to the total flux).
Finally, the stellar radius is consistently well constrained across all scenarios, with a pooled uncertainty of $(2.9 \pm 2.8) \times 10^{-4} R_\odot$.

The bias statistics in \autoref{tab:precision_accuracy} show no concerning systematic offsets.
For all parameters and scenarios, the mean normalized biases are within $1\sigma$ of zero, and the root-mean-square normalized biases are generally below unity, with the largest value reaching $1.2\sigma$.
These results indicate that the inferred stellar parameters are accurate within their quoted uncertainties across the full range of variability scenarios considered.

\FloatBarrier

\section{Discussion}
\label{sec:discussion}

\subsection{Physical Interpretation of Model Preference}
\label{sec:interpretation}

Of the 160 simulated datasets, only eight show a weak preference for the one-component model ($\Delta \ln \mathcal{Z}\approx 3.1{-}3.6$; \autoref{tab:exceptions}).
These cases occur exclusively in the giant-spot scenarios and only at epochs when the true spot filling factor is extremely small, on the order of a few times 10--100\,ppm.
At such low coverages, spots contribute negligibly to the simulated disk-integrated stellar spectrum, rendering the star spectroscopically indistinguishable from a homogeneous photosphere.
Under these conditions, the data do not justify the additional complexity of a two-component model, and the Bayesian model selection correctly favors the simpler description.

This behavior is expected given the achievable measurement precision.
The uncertainty on the spot filling factor depends on the assumed spot morphology, ranging from an average of 0.027\% in the solar-like spot cases to 0.64\% in the giant-spot cases, with a pooled uncertainty across all scenarios of
$\langle \sigma(f_{\mathrm{spot}}) \rangle \approx 0.32\% = 3200$\,ppm (\autoref{tab:precision_accuracy})\vone{.}
For the giant-spot scenarios most relevant here, the average uncertainty is
more than an order of magnitude larger than the true spot coverages in these \vone{eight cases, explaining quantitatively why two-component models were not preferred}.
These results define a practical sensitivity floor for detecting spot signatures in disk-integrated \pandora{} spectra.
While this floor varies from 0.027\% to 0.64\% depending on the spot morphology, we adopt the overall average of 0.32\% (3200\,ppm) as a representative value.

Crucially, this regime is not problematic for \pandora{}'s science goals.
Spot coverages that fall below this sensitivity threshold also have a negligible impact on exoplanet transmission spectra, as discussed in \autoref{sec:corrections}.
The occasional preference for a one-component stellar model therefore reflects an intrinsic detection limit of the data rather than a failure of the retrieval framework.

\vone{\subsection{Impact of Model Fidelity}}
\label{sec:fidelity}

Model fidelity is a foundational assumption of all model-based stellar contamination corrections and has previously been identified as a dominant limitation in transmission spectroscopy \citep{Rackham2023}.
Cross-retrieval experiments---in which simulated data generated from one stellar model grid are fitted using a different one---have shown that mismatches between stellar atmosphere models can lead to biased parameter inferences and, consequently, incorrect contamination corrections \citep{Iyer2020, Rackham2024}.
These effects are particularly pronounced for JWST observations, where broad wavelength coverage, moderate spectral resolution, and high precision amplify discrepancies between models.

Here, we test the impact of model fidelity in the context of \pandora{} observations by repeating our full set of retrievals using a different grid of stellar spectral models.
We use the PHOENIX NewEra model grid \citep{Hauschildt2025}, which incorporates updated atomic and molecular opacities based on expanded line lists, leading to systematic differences in cool-star spectra relative to the PHOENIX ACES grid.
We adopt the low-resolution version of this grid, which spans a wavelength range of 0.25--2.5\,$\micron$ with a sampling of $0.1$\,\AA.
As with our nominal analysis, the spectra are degraded to match the resolution and binning of \pandora{} observations.
All other aspects of the retrieval framework---including the data, noise properties, and model parameterization---are held fixed, isolating the impact of the underlying stellar atmosphere models.

We find that the statistical behavior of the retrievals is largely unchanged under this alternative grid.
In all cases, two-component models are strongly preferred over single-component models, and the inferred parameter uncertainties are comparable to those obtained with our nominal grid.
However, the retrieved parameter values are systematically biased, implying that any stellar contamination corrections derived from these fits would be incorrect.
This result reinforces the conclusion from previous studies \citep{Rackham2024} that model fidelity, rather than statistical precision, is the dominant factor governing the accuracy of model-based corrections.

Critically, these failures are not subtle and would be readily apparent from the quality of the fits.
For our nominal two-component retrievals using the PHOENIX ACES model grid, the median reduced $\chi^2$ is 1.0, indicating that the model provides an adequate description of the data and lending clear credence to the inferred parameters.
In contrast, the corresponding NewEra retrievals yield a median reduced $\chi^2$ of ${\sim}3500$, reflecting a severe mismatch between the model and the data.
Such large $\chi^2$ values, along with structured residuals, provide a clear and quantitative diagnostic that the model is inadequate and that the resulting inferences should not be used for correction.

We therefore emphasize that model-based stellar contamination corrections are conditionally valid:
they can be trusted when the model reproduces the data at the level of the observational uncertainties and rejected otherwise.
Importantly, our results show that when model fidelity is insufficient to support reliable corrections, this failure is accompanied by dramatically elevated $\chi^2$ values, ensuring that such cases can be identified and excluded in practice.
For the remainder of this work, we assume model fidelity by focusing on the results from our nominal retrievals.
Alternative empirical approaches to mitigating stellar contamination \citep[e.g.,][]{Berardo2024, Espinoza2025_T1, Allen2026} offer a complementary path forward but are beyond the scope of this study.

\vone{\subsection{Stellar Activity Cycles}}
\label{sec:cycles}

Stellar activity is time variable, with spot coverage and photometric variability evolving on timescales ranging from stellar rotation periods of a few days to weeks up to multi-year activity cycles. 
Long-baseline photometric monitoring from \kepler{} and \tess{} has demonstrated that at least one \pandora{} target, HAT-P-11, shows significant changes in variability amplitude both between the \kepler{} and \tess{} eras and over the more than 5\,yr of TESS observations \citep{Niraula2026}.
However, the nominal \pandora{} mission lifetime ($\sim$1\,yr), combined with target visibility constraints that typically limit observing campaigns to roughly half of this duration \citep{Foote2023}, implies that \pandora{} alone is unlikely to robustly detect or characterize long-term stellar activity cycles for most targets. 

In this context, the primary opportunity for constraining activity cycles lies in combining \pandora{} observations with longer-baseline photometric datasets. 
As was done for HAT-P-11 \citep{Niraula2026}, similar joint analyses leveraging \pandora{} data alongside archival ground-based, \tess{}, \hst{}, and \jwst{} data may allow \pandora{} observations to be placed within a broader activity-cycle framework, even if the mission itself does not independently resolve such cycles.

\subsection{Stellar Contamination Corrections}
\label{sec:corrections}

Here, we quantify how stellar surface heterogeneity in our simulated scenarios propagates into \pandora{} transmission spectra and assess the extent to which TLS signals can be corrected using stellar parameter inferences alone.
We first define the wavelength-dependent \textit{true} contamination signals for each simulated visit, then compute the corresponding \textit{inferred} contamination signals implied by the stellar retrieval posteriors, and finally evaluate the residual contamination that remains after applying these corrections.

\subsubsection{Definition of True Contamination Signals}

We begin by defining the \textit{true} level of stellar contamination expected in \pandora{} transmission spectra for each simulated visit.
For simplicity, we adopt a wavelength-independent planetary transit depth of $D_0 = 1\%$.

For each 12-hr pre- and post-transit epoch in the simulated time series, we use the true photospheric parameters from the simulation: $T_\mathrm{phot}$, $T_\mathrm{spot}$, and $f_\mathrm{spot}(t)$.
Using the same forward-modeling framework described in \autoref{sec:data}, we generate photosphere and spot spectra ($S_\mathrm{phot}$ and $S_\mathrm{spot}$) and compute the stellar contamination factor
\begin{equation}
\epsilon(\lambda) =
\frac{(1 - f_{\rm chord})\,S_{\rm phot}(\lambda) + f_{\rm chord}\,S_{\rm spot}(\lambda)}
     {(1 - f_{\rm disk})\,S_{\rm phot}(\lambda) + f_{\rm disk}\,S_{\rm spot}(\lambda)},
\label{eq:epsilon}
\end{equation}
where $f_\mathrm{chord}$ is the spot filling factor within the transit chord and $f_\mathrm{disk}$ is the full-disk spot filling factor.
The corresponding imprint of stellar contamination on the observed transit depth is
\begin{equation}
\Delta D(\lambda) = D_0 \,[\epsilon(\lambda) - 1].
\end{equation}

For the giant-spot scenarios, we assume a maximally unocculted spot configuration with $f_\mathrm{chord} = 0$ and $f_\mathrm{disk} = f_\mathrm{spot}$.
This geometry is appropriate for cases in which a small number of large active regions are present on the stellar disk and either do not intersect the transit chord or, if they do, produce spot-crossing events that can be readily identified and modeled.

For the solar-like spot scenarios, the larger spot filling factors lead to a broader range of physically plausible geometries.
We therefore compute contamination signals for two approximately bounding configurations:
a ``maximum'' contamination case identical to the giant-spot prescription ($f_\mathrm{chord} = 0 $ and $f_\mathrm{disk} = f_\mathrm{spot}$), and a ``moderate'' contamination case in which the transit chord samples a representative but not identical fraction of spotted and unspotted regions compared to the stellar disk as a whole.

For this moderate case, we treat the chord-integrated spot filling factor as a stochastic realization of the disk-averaged coverage.
Specifically, we draw $f_\mathrm{chord}$ from a Beta distribution, $f_\mathrm{chord} \sim \mathrm{Beta}(\alpha,\beta)$, with shape parameters $\alpha = \mu / 0.1$ and $\beta = (1-\mu) / 0.1$, where $\mu \equiv f_\mathrm{disk}$.
This prescription yields a distribution with mean $\langle f_\mathrm{chord} \rangle = \mu$ and variance $\mathrm{Var}(f_\mathrm{chord}) = \mu (1 - \mu) \times 0.1$, corresponding to a surface-sampling process in which the transit chord probes $\sim$10\% of the stellar disk, approximately as expected for a planet producing a 1\% transit depth.
For each epoch, we propagate this distribution through \autoref{eq:epsilon} to define a plausible true contamination signal for the moderate solar-like spots scenario.

Analogous to the giant-spot and solar-like spot configurations, these contamination cases are designed to span a plausible range of stellar activity levels that \pandora{} may encounter.
The \pandora{} target list includes systems with a wide diversity of spot morphologies.
The mid-M dwarf TOI-3884, for example, hosts a giant polar spot \citep{Almenara2022, Libby-Roberts2023, Sagynbayeva2025} that persists over years \citep{Chakraborty2025, Mori2025, Tamburo2025} and can significantly impact transmission spectra when left unaccounted \citep{Murray2026}.
At the other extreme, the mid-K dwarf HAT-P-11 has smaller, more numerous spots with solar-like sizes that emerge at preferential latitudes like sunspots \citep{Morris2017}.
Rather than modeling each target individually, the scenarios considered here are intended to bracket this diversity and capture the range of contamination signals expected across the full sample.

\subsubsection{Calculation of Inferred Contamination Signals}

To compute the \textit{inferred} stellar contamination signals, we forward-modeled the posterior distributions from the stellar retrievals through the same contamination formalism used to define the true signals.
For each dataset and epoch, we drew Monte Carlo samples from the posterior distributions of the relevant two-component stellar model parameters: $T_\mathrm{phot}$, $T_\mathrm{spot}$, and $f_\mathrm{spot}$.
For each posterior draw, we evaluated $\epsilon(\lambda)$ using \autoref{eq:epsilon}.

For the giant and ``maximum'' solar-like spots cases, we assumed an unspotted transit chord by setting $f_\mathrm{chord} = 0$ and $f_\mathrm{disk}$ equal to the sampled spot coverage.
For the  ``moderate'' solar-like spots case, we drew $f_\mathrm{chord}$ from the same Beta distribution defined above, ensuring that the inferred contamination signals marginalize over the same range of physically plausible transit geometries as the true signals.

For each epoch and contamination scenario, we summarized the resulting ensemble of contamination spectra by computing the median and central 68\% credible intervals.
For epochs in which the one-component model was preferred, we set the contamination factor to unity ($\epsilon(\lambda) = 1$ by definition), corresponding to zero inferred contamination.

\subsubsection{Residual Contamination after Correction}

\begin{figure*}[htp]
    \centering
    \includegraphics[width=\textwidth]{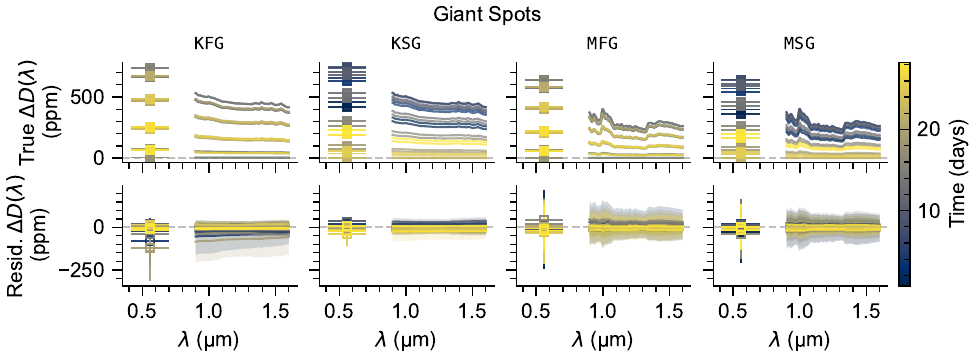}
    \includegraphics[width=\textwidth]{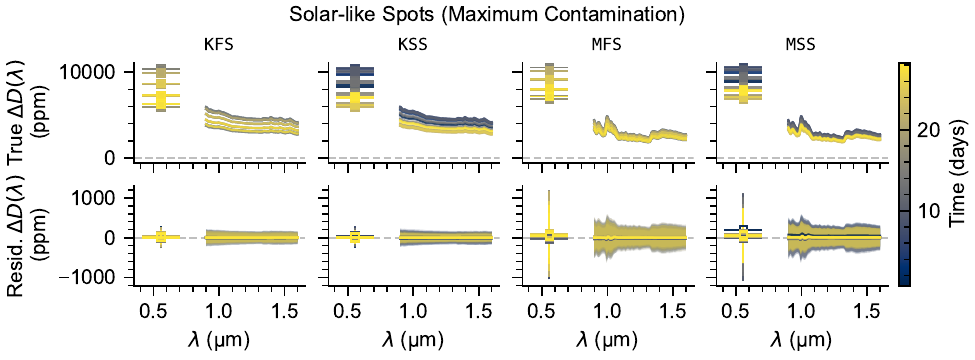}
    \includegraphics[width=\textwidth]{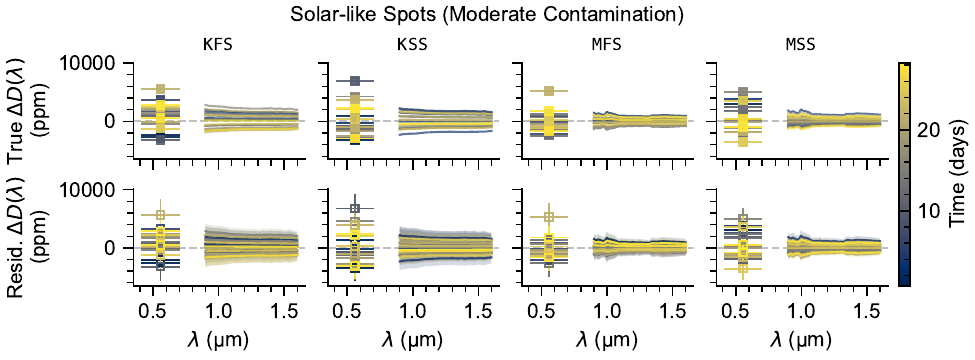}
    \caption{
    True and residual stellar contamination signals for the spot prescriptions we consider: giant spots (top), solar-like spots with maximum contamination (middle), and solar-like spots with moderate contamination (bottom).
    In each row, columns correspond to the four related scenarios (see \autoref{tab:scenarios}).
    Upper subpanels show the true contamination signal, $\Delta D(\lambda)$, across the VISDA bandpass (points with horizontal error bars) and the NIRDA wavelength range (curves).
    All 20 epochs are shown, and colors encode the epoch time.
    Lower subpanels show the residual contamination, $\Delta D_{\rm resid}(\lambda)$, remaining after applying the inferred correction, with shaded regions indicating the propagated uncertainty from the stellar retrieval posteriors.
    }
    \label{fig:contamination_truth_and_residuals}
\end{figure*}

\begin{deluxetable*}{llcccc}[!t]
\tablecaption{True and Residual Stellar Contamination Summary \label{tab:contam_summary}}
\tablehead{
\colhead{Scenario} & \colhead{Case} & \multicolumn{2}{c}{True (ppm)} & \multicolumn{2}{c}{Residual (ppm)} \\
 &  & \colhead{VISDA} & \colhead{NIRDA} & \colhead{VISDA} & \colhead{NIRDA}
}
\startdata
$\texttt{KFG}$ & giant & $370\,\pm\,270$ & $230\,\pm\,170$ & $-10\,\pm\,22$ & $-6\,\pm\,16$ \\
$\texttt{KFS}$ & max. & $8100\,\pm\,1600$ & $4000\,\pm\,620$ & $5.8\,\pm\,17$ & $1.6\,\pm\,9.7$ \\
$\texttt{KFS}$ & mod. & $710\,\pm\,2200$ & $200\,\pm\,1200$ & $720\,\pm\,2100$ & $210\,\pm\,1200$ \\
$\texttt{KSG}$ & giant & $370\,\pm\,270$ & $230\,\pm\,170$ & $-3.9\,\pm\,15$ & $-2.8\,\pm\,14$ \\
$\texttt{KSS}$ & max. & $8100\,\pm\,1600$ & $4000\,\pm\,630$ & $4.4\,\pm\,17$ & $1.4\,\pm\,10$ \\
$\texttt{KSS}$ & mod. & $-26\,\pm\,2700$ & $87\,\pm\,1100$ & $52\,\pm\,2700$ & $120\,\pm\,1100$ \\
$\texttt{MFG}$ & giant & $320\,\pm\,230$ & $140\,\pm\,100$ & $0.79\,\pm\,17$ & $4.9\,\pm\,16$ \\
$\texttt{MFS}$ & max. & $8600\,\pm\,1300$ & $2800\,\pm\,290$ & $11\,\pm\,55$ & $0.76\,\pm\,19$ \\
$\texttt{MFS}$ & mod. & $-210\,\pm\,1700$ & $22\,\pm\,600$ & $-75\,\pm\,1700$ & $64\,\pm\,590$ \\
$\texttt{MSG}$ & giant & $320\,\pm\,230$ & $140\,\pm\,110$ & $-1.6\,\pm\,14$ & $3.3\,\pm\,11$ \\
$\texttt{MSS}$ & max. & $8600\,\pm\,1300$ & $2800\,\pm\,300$ & $34\,\pm\,60$ & $8.3\,\pm\,23$ \\
$\texttt{MSS}$ & mod. & $680\,\pm\,2200$ & $100\,\pm\,650$ & $840\,\pm\,2300$ & $150\,\pm\,650$ \\
\hline
All & giant & $340\,\pm\,250$ & $190\,\pm\,140$ & $-3.7\,\pm\,18$ & $-0.15\,\pm\,15$ \\
All & max. & $8400\,\pm\,1500$ & $3400\,\pm\,770$ & $14\,\pm\,43$ & $3\,\pm\,16$ \\
All & mod. & $290\,\pm\,2200$ & $100\,\pm\,900$ & $380\,\pm\,2200$ & $140\,\pm\,910$ \\
\enddata
\tablecomments{
Values summarize the stellar contamination signal for a 1\% transit depth, expressed in parts per million (ppm), for each simulated stellar scenario and observing case.
The ``True'' columns report the injected contamination amplitude, while the ``Residual'' columns report the remaining contamination after applying the stellar correction procedure.
Results are shown separately for the Pandora VISDA and NIRDA bandpasses.
Quoted values correspond to the mean and standard deviation of the contamination distribution across realizations.
NIRDA values are also averaged over resolution elements.
Scenario labels are defined in \autoref{tab:scenarios}.
Cases labeled ``max.'' and ``mod.'' correspond to maximally and moderately active stellar configurations, respectively.
Rows labeled ``All'' report aggregate statistics across all scenarios within each case.
}
\end{deluxetable*}

We assess the effectiveness of stellar contamination corrections by comparing the true contamination signals defined above to the residual contamination that remains after applying the inferred corrections.
\autoref{fig:contamination_truth_and_residuals} shows the wavelength-dependent true and residual contamination signals for all scenarios, while \autoref{tab:contam_summary} summarizes their amplitudes in the VISDA and NIRDA bandpasses.

For the giant-spot scenarios, the true contamination signals are modest, with typical amplitudes of a few hundred ppm.
Because these cases assume an unocculted transit chord and a small number of large active regions, the spot geometry is well matched to the correction model.
As a result, the inferred corrections removed nearly all of the contamination, reducing the residual signals from the order of ${\sim}100$\,ppm to the $\sim$1--10\,ppm level across both VISDA and NIRDA.
This shows that when the spot geometry is simple, the stellar spectrum contains enough information to correct stellar contamination to well below the precision floor of \pandora{} transmission spectroscopy.

A similar conclusion holds for the solar-like spot scenarios in the ``maximum'' contamination configuration.
Although the true contamination signals in these cases are substantially larger---reaching several thousand ppm---the assumption of a fully unocculted chord is consistent between the true and inferred signals.
Under this matched-geometry assumption, the applied corrections reduce the residual contamination to $\sim$10--50\,ppm, representing a suppression of more than two orders of magnitude.
Even for highly spotted stars, the dominant contamination signal can therefore be mitigated, provided that the assumed spot geometry is appropriate.

The most challenging regime is the solar-like spot scenario with ``moderate'' contamination.
In this case, individual visits show contamination signals of the order of ${\sim}10^3$\,ppm---comparable to, but somewhat smaller than, the maximum case---yet with substantially larger visit-to-visit scatter.
Because the transit chord samples a stochastic realization of the spotted photosphere, both the true and inferred contamination signals fluctuate about zero with large dispersion.
Crucially, the stellar spectrum alone does not uniquely constrain the relative spot filling factor of the transit chord and the full stellar disk.
As a consequence, no single correction can be derived that systematically removes the contamination, and the residual signals closely resemble the true signals in both amplitude and variability.

While this limits the effectiveness of deterministic corrections in the moderate case, it also yields an important diagnostic.
The width of the inferred contamination distribution directly quantifies the systematic noise floor imposed by uncorrected stellar heterogeneity that cannot be corrected using disk-integrated stellar spectra alone.
In this sense, an analysis of the stellar spectrum that reveals a large spot filling factor can alert us to a noise source from stellar contamination that must be accounted for in some other way unless the distribution of spots can be constrained more directly.

The broader implication is that when stellar spectra indicate large spot filling factors, care must be taken in assuming how those spots are distributed relative to the transit chord.
Corrections derived solely from disk-integrated stellar spectra may be incomplete or misleading in such cases.
Additional constraints---such as spot-crossing events or chromatic stellar contamination signatures within the transmission spectrum itself---can, in principle, break this degeneracy.
In particular, these simulations show that the wavelength dependence of stellar contamination should be readily identifiable in the \pandora{} transmission spectrum in such regimes, enabling joint constraints on stellar and planetary signals from analyses of in-transit data.
Exploring these additional possible constraints is beyond the scope of this work, but the results presented here provide a quantitative framework for understanding when stellar contamination can be corrected from stellar spectra alone, when it cannot, and how the associated uncertainty should be incorporated into exoplanet atmosphere studies.

\subsection{Implications for Pandora Science Goals}

An important lesson from these simulations is that photometric variability amplitude alone is an ambiguous indicator of stellar contamination and its correctability \citep[see also][]{Rackham2018, Rackham2019}.
All scenarios considered here were constructed to exhibit a 1\% variability amplitude, yet they span orders of magnitude in spot filling factor, depending on the assumed spot size.
They also produce significantly different contamination signals in transmission spectra depending on the spot distribution.
Variability levels observed by \pandora{} should therefore be interpreted in the context of inferred spot distributions rather than as a direct proxy for contamination severity.

The giant-spot scenarios show that, for some physically plausible spot distributions, disk-integrated stellar spectra are sufficient to characterize stellar heterogeneity and to derive effective contamination corrections.
In these cases, a small number of large, active regions produce modest spot filling factors and contamination signals that are well constrained by the stellar spectrum, allowing corrections that suppress stellar contamination to below the expected precision of \pandora{} transmission spectra \citep[30--100\,ppm;][]{Rotman2026}.
For such targets, \pandora{}'s stellar observations can fully mitigate stellar contamination without requiring additional assumptions about spot distributions.

By contrast, the solar-like spot scenarios illustrate regimes in which stellar spectra alone are insufficient to derive reliable contamination corrections.
When spot filling factors are high, both the maximum and moderate contamination cases show that the relationship between the disk-integrated and transit-chord spot coverages becomes a dominant source of uncertainty.
In these regimes, deterministic corrections based solely on stellar spectra may be incomplete or misleading, even when the stellar parameters themselves are well constrained.
Importantly, \pandora{}'s out-of-transit observations play a critical diagnostic role in identifying these challenging regimes, thereby distinguishing cases in which stellar contamination can be robustly corrected via stellar spectra alone from those in which additional constraints---such as spot-crossing events or joint stellar and planetary retrievals---are required to fully assess stellar contamination in \pandora{} transmission spectra.
This ability to classify targets via stellar observations represents a positive and mission-enabling outcome of this methodology.

\subsection{Limitations and Future Extensions}

The analysis presented here adopts a deliberately simplified framework to enable a focused and quantitative assessment of \pandora{}'s stellar contamination capabilities.
We have intentionally designed the investigation to explore outcomes for a feasible range of spot filling factors at \pandora{}'s wavelength coverage, cadence, and precision, rather than to model the full complexity of stellar surfaces.
Here, we outline 
\vone{two}
key limitations of our current framework, justify these design choices, and explore avenues for future extensions.

\subsubsection{Neglecting Faculae}

We consider only two photospheric components (quiet photosphere and spots) and do not model the effects of faculae.
This choice is 
motivated both by limitations in available stellar atmosphere models and by the expected wavelength dependence of facular signals.


First, realistic facular spectra remain poorly constrained.
Existing studies \citep[e.g.,][]{Norris2017, Norris2023, Witzke2022} show that commonly adopted approximations---such as treating faculae as spectra of hotter photospheres---can introduce systematic errors and that facular contrasts depend sensitively on viewing angle and magnetic geometry.
These uncertainties are particularly acute for K and M dwarfs, where faculae may exhibit reduced contrast or even appear dark in some regimes \citep[][A.\,I.~Shapiro et al.\ 2026, in preparation]{Witzke2022, Norris2023}.

Observational constraints on facular covering fractions for active K and M dwarfs are similarly uncertain, with estimates ranging from ${\sim}5\%$ to ${\gtrsim}60\%$ depending on model assumptions \citep{Rackham2018, Rackham2019}.
By contrast, the Sun exhibits a maximum facular coverage of only ${\sim}3\%$ \citep{shapiro2014}.
Empirically, evidence for bright heterogeneities---interpreted as faculae---covering stellar disks at the few-percent level have been identified from active region crossings \citep{Espinoza2019} and retrievals of transmission spectra \citep[e.g.,][]{Rackham2017, Kirk2021} in the optical.
However, NIR spectra of active K dwarfs that favor two-component (spot and photosphere) models do not show a preference for an additional facular component \citep{Narrett2024, Niraula2026}.

Despite these uncertainties, faculae could in principle bias the inferred photospheric properties from VISDA photometry and thereby propagate into the inferred stellar contamination correction in the NIR.
To assess the magnitude of this effect, we constructed a simple toy model based on the same framework used throughout this work.
We approximated faculae as a hotter component with $T_\mathrm{fac} = T_\mathrm{phot} + 100$\,K \citep[following][]{Rackham2018, Rackham2019} and evaluated their impact on both VISDA and NIRDA observables for our example K and M dwarfs.
We considered facular filling factors of 10--20\%, consistent with moderately active stars.

Using this model, we find that faculae induce fractional flux changes of $\sim$1.4--2.7\% in VISDA and $\sim$0.6--1.1\% in NIRDA for K dwarfs, and $\sim$3.2--6.3\% (VISDA) versus $\sim$1.0--2.0\% (NIRDA) for M dwarfs.
Thus, facular signals are consistently suppressed in the NIR by factors of $\sim$2--3 relative to the visible.
These values are likely conservative:
for example, the active K dwarf HAT-P-11 shows variability of only $\sim$0.35\% in the WFC3/G102 bandpass, which lies between VISDA and NIRDA, suggesting that our assumed contrasts may overestimate the true effect.

For comparison, the average spot filling factors explored in our simulations (\autoref{tab:spot_coverage}, \autoref{fig:fspot}) produce substantially larger signals.
For K dwarfs with giant spots ($\langle f_\mathrm{spot} \rangle = 0.04$), spots induce flux changes of 3.5\% in VISDA and 2.2\% in NIRDA.
For the solar-like spot distributions ($\langle f_\mathrm{spot} \rangle = 0.51$), these increase dramatically to 44\% (VISDA) and 28\% (NIRDA).
A similar trend is seen for M dwarfs:
giant-spot cases ($\langle f_\mathrm{spot} \rangle = 0.04$) yield 3.0\% (VISDA) and 1.4\% (NIRDA), while solar-like spot cases ($\langle f_\mathrm{spot} \rangle = 0.61$) produce 46\% (VISDA) and 22\% (NIRDA) flux changes.
Even in the most modest (giant-spot) scenarios, spot-induced variability is comparable to or larger than the facular signals estimated above, while in the solar-like spot regimes it exceeds them by more than an order of magnitude.
Thus, across all cases considered, spots dominate the integrated spectral variability, particularly in the NIR where our constraints are strongest.

Taken together, these results provide a simple demonstration that while faculae can influence the visible-band flux---and thus may bias inferred filling factors at some level---their direct contribution to the NIR spectrum is comparatively small.
Any impact of facular-induced biases from VISDA onto the NIR contamination correction is therefore expected to be secondary to the effects of spots.
In cases where faculae introduce additional complexity in \pandora{} observations, a practical mitigation strategy is to place greater weight on NIRDA observations, where their impact is intrinsically reduced.
We emphasize that a more complete treatment of faculae will ultimately require improved magnetohydrodynamic stellar atmosphere models and empirical constraints, particularly for cool stars.



\subsubsection{Intravisit Timing Assumptions}

In our nominal simulations, we assume that each transit is centered within a visit that includes symmetric pre- and post-transit baselines, and that the 10 epochs are obtained consecutively.
In practice, scheduling constraints \citep{Foote2023} may lead to asymmetric baseline coverage (e.g., more pre- than post-transit time) and epochs that are separated in time rather than acquired back-to-back.

Our inference framework depends primarily on the total out-of-transit baseline duration within each visit rather than on its symmetry about transit.
Moderate asymmetries in pre- and post-transit coverage are therefore unlikely to qualitatively affect our conclusions.
Non-consecutive sampling, however, would probe different rotational phases and potentially different active-region configurations, which could yield a broader range of spot filling factors across epochs.
Nonetheless, our simulations are intentionally designed to explore a wide range of spot filling factors
\vone{(0--68\%)}, 
thereby capturing the variability expected from sampling at disparate stellar phases.
Thus, although we do not explicitly model long-term evolution between widely separated visits, we do not expect non-consecutive sampling to alter our conclusions.

\subsubsection{Independent Treatment of Epochs}

Each simulated epoch in our analysis is modeled independently, without enforcing shared stellar parameters or spectral components across visits.
Rather than performing a joint, multi-epoch retrieval, we analyze each visit separately.
In principle, a joint fit could strengthen constraints on stellar properties by leveraging temporal coherence across epochs.
For example, it could help determine whether cases favoring a one-component model reflect intrinsically low spot filling factors or instead arise from statistical fluctuations.

Such an approach, however, introduces additional assumptions\vone{ and modeling complexity.}
\vone{In particular, it would require assumptions about the temporal stability of the underlying spectral components.}
In reality, spot spectra can change as active regions evolve and the relative contributions of umbral and penumbral components vary \citep{Solanki2003, Berdyugina2005}.
By treating epochs independently, our framework naturally accommodates this spectral evolution without imposing restrictive priors.

Additionally, a multi-epoch approach would require modeling the evolution of spot filling factors across epochs, either by introducing a large number of free parameters in the fit or enforcing a specific temporal relationship.
In the current framework, we instead treat the spot filling factor at each visit independently.
The simulated sinusoidal variability is used only to define the instantaneous filling factor at each epoch and is not imposed in the retrieval, as doing so would be overly restrictive.

Gaussian processes with quasiperiodic kernels \citep{Rasmussen2006, Aigrain2023} provide a flexible and efficient framework for modeling time-dependent spot filling factors in multi-epoch fits.
Such models capture quasiperiodic behavior without imposing rigid functional forms or largely increasing the number of free parameters.
We are exploring the use of this approach for future multi-epoch analyses.


\section{Conclusions}
\label{sec:conclusions}

In this work, we used a suite of end-to-end simulations to assess the Pandora SmallSat Mission's ability to characterize stellar photospheres and to correct for stellar contamination in exoplanet transmission spectra using out-of-transit observations alone.
We constructed eight representative K- and M-dwarf scenarios spanning rotation rate and spot morphology, all tuned to exhibit the same 1\% visible-band variability amplitude.
For each case, we generated realistic, time-dependent stellar spectra, propagated them through a detailed \pandora{} instrument and noise model, and performed independent stellar retrievals on a total of 160 simulated datasets.
We then forward-modeled the inferred stellar parameters to quantify true, inferred, and residual stellar contamination signals under a range of physically motivated spot geometries.

Our main findings are as follows.

\begin{enumerate}
    \item Under the simulated scenarios considered here, \pandora{} can recover stellar photospheric properties with high precision and accuracy across a broad range of stellar activity regimes (\autoref{fig:inferences_K}, \autoref{fig:inferences_M}, \autoref{tab:precision_accuracy}).
    Photospheric temperatures are constrained to ${\approx}30$\,K for both K and M dwarfs.
    The stellar radius, which functions as a flux-scaling parameter in our fits under the assumption of fixed stellar distance, is recovered with negligible bias at the ${<}0.1\%$ level.
    Spot temperatures and filling factors are accurately inferred whenever spots contribute appreciably to the disk-integrated spectrum, with uncertainties that scale naturally with the true spot coverage.
    These results show that \pandora{}'s combined visible and NIR observations provide sufficient information to robustly characterize the heterogeneous stellar photospheres of its active K- and M-dwarf targets in most astrophysically relevant regimes.
    
    \item Given sufficiently accurate models, Bayesian model selection defines a physically meaningful sensitivity floor for detecting stellar heterogeneity.
    Two-component stellar spectral models are strongly preferred in 95\% of the simulated datasets.
    The remaining cases (\autoref{tab:exceptions}) favor a one-component model only when the true spot filling factor falls well below the typical retrieval uncertainty ($0.32\% = 3200$\,ppm).
    This behavior reflects an intrinsic detection limit rather than a failure of the retrieval framework and establishes a practical floor of order 0.3\% in spot coverage below which stellar heterogeneity is neither detectable in \pandora{} spectra nor expected to significantly impact transmission spectroscopy.

    \item Photometric variability amplitude alone is not a reliable indicator of stellar contamination severity or correctability.
    Although all simulated stars exhibit the same 1\% variability amplitude in the visible band, they span more than an order of magnitude of spot filling factor (\autoref{tab:spot_coverage}) and produce contamination signals ranging from a ${\sim}100$\,ppm to ${>}10^4$\,ppm (\autoref{fig:contamination_truth_and_residuals}, \autoref{tab:contam_summary}; see also \citealt{Rackham2018, Rackham2019}).
    The degree of stellar contamination depends primarily on spot morphology and surface distribution rather than variability amplitude itself, underscoring the need for spectroscopic characterization of stellar photospheres rather than reliance on photometric inferences alone.

    \item When the transit chord can be rightfully treated as unspotted and spectral components are well modeled, time-averaged \pandora{} spectrophotometry can be used to correct contamination signals to well below the expected precision of \pandora{} transmission spectra.
    In giant-spot scenarios and in solar-like spot scenarios with fully unspotted transit chords, true contamination signals ranging from order $10^2$\,ppm to order $10^3$\,ppm are suppressed by more than two orders of magnitude after correction (\autoref{tab:contam_summary}).
    The resulting residual contamination levels, typically ${\lesssim}10$\,ppm across the \pandora{} bandpass, lie comfortably below the anticipated 30--100\,ppm precision of \pandora{} transmission spectra \citep{Rotman2026}.
    In these regimes, disk-integrated stellar spectra contain sufficient information to define a correction to fully mitigate stellar contamination.

    \item For complex spot geometries and distributions, stellar spectra alone cannot uniquely determine stellar contamination.
    In solar-like spot scenarios where the transit chord samples a stochastic subset of the stellar surface, the relationship between disk-averaged and chord-averaged spot coverage remains unconstrained.
    In this regime, residual contamination remains at the ${\gtrsim}10^3$\,ppm level (\autoref{fig:contamination_truth_and_residuals}, \autoref{tab:contam_summary})---comparable to the true contamination signal---and deterministic corrections derived from stellar spectra alone are ineffective.
    Crucially, these cases reflect an intrinsic geometric degeneracy for highly spotted stars rather than insufficient data quality, and the analysis presented here shows that \pandora{} stellar observations can aid in identifying such cases.
    This capability highlights the value of \pandora{} for both enabling effective direct correction when possible and also identifying regimes where complementary constraints, such as spot-crossing events or joint star-and-planet retrievals of transmission spectra, are necessary to fully disentangle stellar and planetary signals.
\end{enumerate}

Looking ahead, these results establish a quantitative framework for interpreting \pandora{} exoplanet observations in the context of \pandora{}'s stellar constraints.
By identifying the regimes in which stellar contamination can be robustly corrected with stellar observations---and those in which more constraints are needed---our results clarify both the power and limitations of time-resolved spectrophotometry.
Future extensions of this work will explore joint multi-epoch and in-transit retrievals, incorporate additional sources of stellar heterogeneity such as faculae, and assess how complementary observables can further break degeneracies between signals from stellar surfaces and 
planetary atmospheres.
Together, these efforts will ensure that \pandora{} transmission spectra can be fully leveraged to study the atmospheres of worlds transiting active cool stars.

\begin{acknowledgments}
\pandora{} is supported by NASA’s Astrophysics Pioneers Program. 
This material is based upon work supported by NASA under award No.\ 80NSSC24K0197.
This material is partly based upon work supported by the National Aeronautics and Space Administration under Agreement No.\ 80NSSC21K0593 for the program ``Alien Earths.''
The results reported herein benefited from collaborations and/or information exchange within NASA’s Nexus for Exoplanet System Science (NExSS) research coordination network sponsored by NASA’s Science Mission Directorate.
This material is based upon work supported by the European Research Council (ERC) Synergy Grant under the European Union’s Horizon 2020 research and innovation program (grant No.\ 101118581---project REVEAL).
\vone{We thank the reviewer for their thoughtful and constructive assessment of the manuscript.}
A.R.I. and R.H. would like to acknowledge support from the NASA Postdoctoral Program by an appointment at the NASA Goddard Space Flight Center, administered by Oak Ridge Associated Universities under contract with NASA.
This publication was partly funded by the Heising-Simons Foundation through grant \#2024-5688.
This work was partly performed under the auspices of the U.S. Department of Energy by the Lawrence Livermore National Laboratory under Contract DE-AC52-07NA27344. The document number is LLNL-JRNL-2016713.
The authors thank the Canadian Space Agency for support through the ROSS Program.
The authors would like to acknowledge the use of OpenAI's ChatGPT in editing portions of this manuscript to improve the clarity of the writing.
\end{acknowledgments}

\begin{contribution}



B.V.R. led the conceptualization of the study, the generation and analysis of the simulated data, and the writing of the manuscript.
A.R.I. and DA contributed to the study design and provided guidance on the scientific scope and interpretation.
P.M., Y.R., K.D.C., B.M.M., and E.A.G. provided valuable feedback on the analysis and manuscript.
All authors contributed to enabling NASA's Pandora SmallSat Mission and to shaping the broader scientific direction of this work.

\end{contribution}

%
\facilities{\pandora{}}

\software{\texttt{Astropy} \citep{astropy:2013, astropy:2018, astropy:2022}; 
          \texttt{ExoCTK} \citep{Bourque2021, Bourque2022};
          \texttt{Matplotlib} \citep{Hunter2007};
          \texttt{NumPy} \citep{numpy};
          \texttt{pandas} \citep{pandas};
          \texttt{pandora-sat} \citep{Hedges2024};
          \texttt{SciPy} \citep{SciPy};
          \texttt{spotter} \citep{Garcia2025};
          \texttt{speclib} \citep{speclib2023, speclib2025};
          \texttt{UltraNest} \citep{Buchner2021}
          }




\bibliography{bibliography, local}{}
\bibliographystyle{aasjournalv7}



\end{document}